\documentclass[12pt,draftcls,onecolumn,english]{IEEEtran}
\usepackage[english]{babel}
\usepackage{cite}
\usepackage{amssymb, amsmath}
\usepackage{graphicx}
\usepackage{color}
\usepackage{url}
\usepackage{multirow}
\usepackage{subfigure}
\usepackage{dblfloatfix}

\usepackage[utf8]{inputenc} 
\usepackage{cite}
\usepackage{amsmath,amssymb,amsfonts}
\usepackage{algorithm,algorithmic}
\usepackage{graphicx}
\usepackage{textcomp}
\usepackage{multirow}
\usepackage{subfigure}

\usepackage{setspace}

\DeclareMathOperator*{\argmin}{arg\,min}

\begin{document}
\title{Mixed Gibbs Sampling Detector in High-Order Modulation Large-Scale MIMO Systems}

\author{{Alex Mussi,}  \, {Taufik Abrão}
\thanks{ T. Abrão is with the Electrical Engineering Department, State University of Londrina, PR, Brazil.  \\
Rod. Celso Garcia Cid - PR445; 86057-970. E-mail: {taufik@uel.br}}	
\thanks{A. Mussi is with Federal Institute of Paraná, Cívica Av. 475. CEP: 85935-000. Assis Chateaubriand. Brazil. E-mail: alex.mussi@ifpr.edu.br}
}

\maketitle

\begin{abstract}
A neighborhood restricted Mixed Gibbs Sampling (MGS) based approach is proposed for low-complexity high-order modulation large-scale Multiple-Input Multiple-Output (LS-MIMO) detection. The proposed LS-MIMO detector applies a neighborhood limitation (NL) on the noisy solution from the MGS at a distance $d$ -- thus, named $d$-simplified MGS ($d$-sMGS) -- in order to mitigate its impact, which can be harmful when a high order modulation is considered. Numerical simulation results considering $64$-QAM demonstrated that the proposed detection method can substantially improve the MGS algorithm convergence, whereas no extra computational complexity per iteration is required. The proposed $d$-sMGS{-based detector suitable for high-order modulation LS-MIMO} further exhibits {improved performance $\times$ complexity tradeoff} when the system loading is high, {i.e.},  {when  $\frac{K}{N}\geq 0.75$}. Also, with increasing the number of dimensions, i.e., increasing number of antennas and/or modulation order, a smaller restriction of $2$-sMGS was shown to be a more interesting choice than $1$-sMGS.
\end{abstract}

\section{Introduction} \label{sec:introduction}
{In order to meet the demands of high transmission capacity, high reliability and spectral and energy efficiency requirements of modern wireless communication systems, the multiple input and output (MIMO) technique has been proposed and considered an appropriate solution due to to their ability to provide multiplexing and diversity gains without the need for additional spectral features. These advantages are further enhanced by large-scale use, called Large-Scale MIMO (LS-MIMO), which has important application in fifth generation (5G) wireless communications. Such structures hold the same benefits as conventional MIMO, however on a larger scale. More properly, LS-MIMO is defined as a transmission/reception design using typically several tens or even hundreds of antennas in at least one of the communication terminals, usually in the base station (BS) \cite{Hoydis2013, Rusek2013}. This turns out to be convenient for the systems in question, since the reduced dimensions of {user equipments (UEs)} {suggest} a single antenna arrangement in each UE; on the other hand, a huge amount of antennas need to be is installed in each BS}.

{However, the LS-MIMO high capacity/spectral efficiency comes with a price: as the number of antennas at BS increases, the computational complexity of data detection {tends} to grow proportionally. Hence, efficient and low-complexity symbol detection techniques becomes critical as the processing of large numbers of signals can become a system bottleneck. It is well known that maximum likelihood (ML) detection could provide optimum symbol detection, but its high complexity forbids it from a practical implementation for MIMO systems. Therefore, sub-optimal linear and non-linear detectors with low complexity are often employed. Many low-complexity LS-MIMO detectors have been proposed in recent literature, including detectors based on a) {\it local neighborhood search}, such as likelihood ascent search (LAS) algorithm \cite{Vardhan2008}, and reactive tabu search (RTS) algorithm \cite{Rajan2009}; b) message passing (MP) algorithms, based on {\it belief propagation} (BP) technique, such that LS-detectors inspired in graphical models, as {\it factor graph} (FG) \cite{Som6-8Jan.2010} and {\it Markov random fields} (MRF) \cite{SuneelJune282009-July32009}; c) minimum mean square error (MMSE) approximation techniques \cite{TangMarc2016, Thanos2017}, which result in low-complexity at the price of achieving good performance only at low system loading factor; d) Markov Chain Monte Carlo (MCMC) techniques, which are based on Gibbs Sampling (GS) \cite{Martino.2018} and its variations \cite{DattaSept.2013,ChoiFeb.2016,Chen2002,Farhang-BoroujenyMay2006}, {emerging as a promising approach to deal} with LS-MIMO structures, since {such techniques} demonstrate a near-optimum performance while require a low-moderate complexity (quadratic order) and also presenting a simple {and} effective way to solve the large-scale detection problem.} 

{From the GS based techniques, in \cite{DattaSept.2013} a strategy of mixing between the conventional GS solution and a random or noisy solution was proposed, which is controlled by a mixing ratio parameter and is called Mixed GS (MGS). The MGS has been shown to solve the stalling problem of the GS detector in low order of modulation, i.e., $4$-QAM. With the modulation order increasement, the multiple restarts (MR) technique is proposed, which restarts the algorithm with a new initial solution, taking advantage of the random evolution of the algorithm and can result in a better cost solution. The MGS-MR detector showed near-optimal performance in $16$-QAM modulation, however, in high modulation order the noisy solution interferes with the convergence of the algorithm, requiring an extra strategy to avoid the impact of this solution. In \cite{Mussi2018} is proposed the use of multiple samples, called averaged MGS (aMSG), in order to minimize this impact, besides a simplification in the target distribution function. Numerical results demonstrate a convergence improvement in high order modulation and high system loading, on the other hand, the choice of sample amount and mixing ratio tends to be difficult. In the present work, a strategy for reducing the solution is also addressed, through a limitation in the neighborhood of the random solution, which presented superior performance to the aMGS, with marginally similar computational complexity.}

{Also related to the MGS detector, in \cite{Hassibi2014} an optimization on mixing time was introduced to accelerate the finding of the optimal solution. Numerical results demonstrated that a mixing time dynamic choice based on SNR can improve convergence, although the stalling problem persisted when a fixed mixing time is adopted. Besides that, these results did not considered the performance behavior in high-order modulation systems. A QR decomposition approach within the MCMC detector was addressed in \cite{Yang2016a,Mandloi2017}, which demonstrated to reduce the number of operations due to the lower triangular matrix feature. Furthermore, based on the concept of multiple random parallel Markov chains, work in \cite{Gao2016} proposes a MR strategy through parallel chains; such strategy reduced the algorithm's running time compared to MGS-MR, despite the increasing of the number of real operations per symbol.}

{The {\it contribution} of this work follows: {\bf i}) a neighborhood limitation (NL) strategy is proposed aiming at improving the MGS convergence rate operating under higher-order modulation and large scale MIMO regime. The proposed strategy, called $d$-sMGS ($d$-simplified MGS), performs a NL in the random solution coming from the mixture used by the MGS detector. As a result, the impact caused by this noisy solution is mitigated and the convergence is increased.} 
{{\bf ii}) an analysis of the performance $\times$ complexity tradeoff is carried out among the proposed $d$-sMGS, the conventional MGS \cite{DattaSept.2013} and the aMGS (averaged MGS) \cite{Mussi2018}, which the latter is an approach that also aims to alleviate the impact caused by the random solution, although the procedure is based on multiple sampling (MS) strategy, which samples the estimated symbol multiple times and performs a mean operation to obtain the result.}

The remainder of this paper is organized as follows. Section \ref{sec:systemodel} presents the adopted large-scale MIMO system model. A review on the MGS technique is presented in section \ref{sec:MGS} {and the MGS based approaches with noisy solution reduced impact are discussed in section \ref{sec:noisy_appr}, while the aMGS approach is described in subsection \ref{subsec:amgs} and the proposed simplified MGS with NL detector for LS-MIMO is developed in {subsection \ref{subsec:amgs}}}. {Computational complexity are presented in section \ref{sec:complexity} and} extensive numerical simulation results are analyzed in section \ref{sec:results}. Conclusion remarks are provided in section \ref{sec:concl}.

\section{System model and problem formulation}\label{sec:systemodel}
\begin{figure}[!htbp]
	\centering
	\includegraphics[width=.65\textwidth]{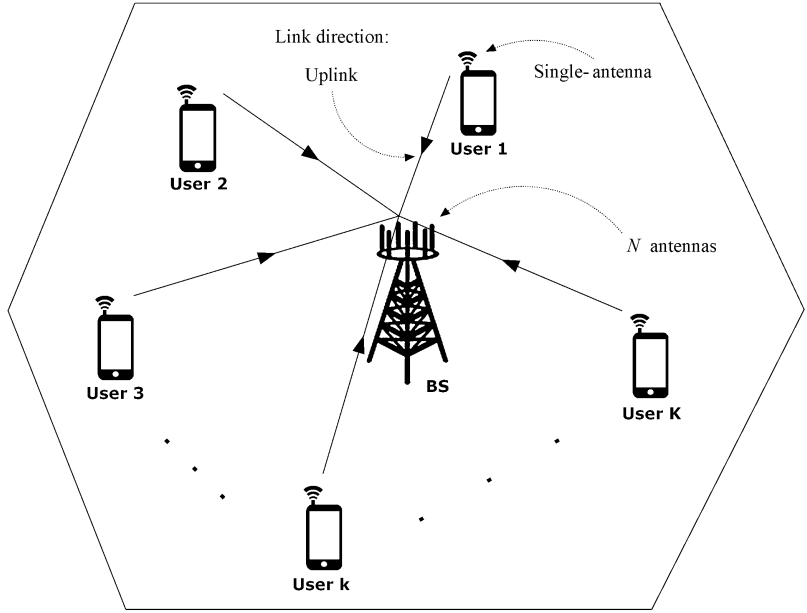}
	\caption{Single-cell uplink LS-MIMO communication system with $K$ single-antenna unit equipments (UEs) and a base-station (BS) with $N$ receive antennas.}
	\label{fig:MIMO_syst}
\end{figure}

We consider an uplink (UL) single-cell MIMO communication system operating in multiplexing gain mode with $K$ active single-antenna users and $N$ receive antennas at the base station (BS), as disposed Fig. \ref{fig:MIMO_syst}. We mainly investigate the performance $\times$ complexity tradeoff of suitable LS-MIMO detection schemes and, for simplicity, the availability of the channel state information at the BS is considered, which also aims to reach the pure efficiency of each detection technique. Thus, the pilot training stage and the respective pilot contamination effect have not taken into account in such context.

Moreover, for simplicity, the communication channel is assumed to be frequency-flat fading, compound by the complex channel matrix $\mathbf{H}_c \in \mathbb{C}^{N \times K}$. The elements of $\mathbf{H}_c$ are all independent complex Gaussian random variables with zero mean and unit variance, i.e., {$H_{c_{i,k}} \sim \mathcal{CN}[0; 1]$}{, where $H_{c_{i,k}}$ denotes the element in the $i$-th row and $k$-th column of the matrix $\mathbf{H}_c$}. Let $\mathbf{s}_c$ be the $K \times 1$ complex vector corresponding to the {$K$ symbols $M$-QAM} transmitted over the single-antenna users, ${\mathbf{s}_c} \in \mathbb{A}_c^{K}$ where $\mathbb{A}_c$ denotes the QAM constellation adopted. The UL received signal, $y_{c_i}$, at the $i$-th BS antenna can be written as:
\begin{eqnarray}
y_{c_i} &=& \sum\limits_{j=1}^{K} H_{c_{i,j}} s_{c_j} + \eta_{c_i},  \qquad {i=1,\ldots,N}  \\
&=& \underbrace{H_{c_{i,k}} s_{c_k}}_{\text{desired signal}}  + \underbrace{\sum\limits_{j=1,j\neq k}^{K} H_{c_{i,j}} s_{c_j} }_{\text{intracellular interference}} + \underbrace{\eta_{c_i}}_{\text{AWGN}}, \nonumber
\end{eqnarray}
{where $y_{c_i}$ denotes the $i$-th element of the complex received signal vector $\mathbf{y}_{c}$ and $s_{c_j}$ is the $j$-th element of $\mathbf{s}_c$.}
{In matrix form, the received signal vector at the BS is re-written as}
{
\begin{equation}\label{eq:rx_signal}
\hspace{0.27\linewidth}{\mathbf{ y}_c  = \mathbf{ H}_c \mathbf{ s}_c} +   {\boldsymbol \eta}_c,
\end{equation}}
where ${\boldsymbol \eta}_{{c}}$ denotes the additive white Gaussian noise (AWGN) {vector,} assumed to be a complex Gaussian random variable with zero mean and variance given by {$\mathbb{E} [ {\boldsymbol \eta}_c {\boldsymbol \eta}_c^H ] = \sigma^2 \mathbf{I}_N$,} where $\sigma^2$ is the noise variance at each receive antenna. 

The average received SNR at each receive antenna can be modelled as { $\gamma = \frac{K P_{\mathrm{s}}}{\sigma^2}$,} where ${P_{\mathrm{s}}}$ is the power of the received symbols. For simplicity, it is considered that the large-scale fading effect has been compensated in such a way that all $K$ users' signals are received with equal power at the BS, and assumed equal to $K P_{\mathrm{s}}$, denoting the total sum power available at the transmitters  \cite{Chockalingam2014}.

In this work, a real-valued system model corresponding to \eqref{eq:rx_signal} is adopted, which is given by:
{
\begin{equation}\label{eq:real_rx_signal}
\hspace{0.31\linewidth}\mathbf{ y}  = \mathbf{ H} \mathbf{ s} +   {\boldsymbol \eta},
\end{equation}}
where $\mathbf{ y} \in \mathbb{R}^{2N \times 1}$, $\mathbf{ H} \in \mathbb{R}^{2N \times 2K}$, $\mathbf{ s} \in \mathbb{R}^{2K \times 1}$, ${\boldsymbol \eta} \in \mathbb{R}^{2N \times 1}${, and defined as}:
\begin{eqnarray}
\mathbf{H} &=& \left[ \begin{matrix}
\mathcal{R}\left(\mathbf{ H}_c\right) \, -\mathcal{I}\left(\mathbf{ H}_c\right) \\
\mathcal{I}\left(\mathbf{ H}_c\right) \quad \mathcal{R}\left(\mathbf{ H}_c\right)
\end{matrix} \right] \\
\mathbf{s} &=& \left[ \begin{matrix}
\mathcal{R}\left(\mathbf{ s}_c\right) \\
\mathcal{I}\left(\mathbf{ s}_c\right) 
\end{matrix} \right], \nonumber \quad
{\boldsymbol \eta} = \left[ \begin{matrix}
\mathcal{R}\left({\boldsymbol \eta}_c\right) \\
\mathcal{I}\left({\boldsymbol \eta}_c\right) 
\end{matrix} \right], \quad
\mathbf{y} = \left[ \begin{matrix}
\mathcal{R}\left(\mathbf{ y}_c\right) \\
\mathcal{I}\left(\mathbf{ y}_c\right) 
\end{matrix} \right]. \nonumber 
\end{eqnarray}

For the QAM complex alphabet $\mathbb{A}_c$, the elements of $\mathbf{s}$  assume {integer} values from the underlying pulse-amplitude modulation (PAM) alphabet $\mathbb{A}$, i.e., $\mathbf{s} \in \mathbb{A}^{2K}$. 

The {\it maximum-likelihood} (ML) {decision rule} is given by{: $\mathbf{s}_{\mathrm{ML}} = \argmin_{\mathbf{\hat{s}} \in \mathbb{A}^{2K}} ||\mathbf{y} - \mathbf{H}\mathbf{\hat{s}}||^2$.} {However}, the ML detector is exponentially complex in $K$, being prohibitive for large $K\cdot N$, which is the case of LS-MIMO systems \cite{Chockalingam2014}.

\section{Conventional method: review of Mixed Gibbs Sampling detection}\label{sec:MGS}
The LS-MIMO detector Mixed Gibbs Sampling (MGS) proposed in \cite{DattaSept.2013} is revisited in this subsection, which is based on the motivation to solve the stalling problem presented in the conventional GS detector.

To sample the estimated symbol at each position, a target distribution \cite{HansenNov2009} is evaluated, which is given by:
\begin{equation}\label{eq:APP_MGS}
p(\hat{s}_{1},\hat{s}_{2},\dots,\hat{s}_{2K}|\mathbf{y}, \mathbf{H}) \propto \exp \left(- \frac{||\mathbf{y} - \mathbf{H} \mathbf{s} ||^2}{\alpha^2 \sigma^2}\right),
\end{equation}
where $\hat{s}_{i}$ denotes the $i$-th position of the estimated symbols vector $\hat{\mathbf{s}}$, $\alpha$ denotes a positive parameter, which tunes the mixing time of the Markov chain \cite{HansenNov2009} and is also called as temperature. The conventional Gibbs sampling detector does not include the $\alpha$ parameter in its sample process, and thus can be viewed as a special case when $\alpha=1$. A larger temperature speeds up the mixing and aims to reduce the higher moments of the number of iterations when finding the correct solution. However, as stated in \cite{DattaSept.2013}, the stalling problem persists even with large $\alpha$.

The MGS detector utilizes a mixing of: {\bf a}) Conventional Gibbs sampling ({i.e.}, $\alpha=1$); and {\bf b}) the infinite temperature version of \eqref{eq:APP_MGS} (i.e., $\alpha=\infty$), resulting in a random and uniform sample from all the possibilities, called a noisy or random solution in this paper. In this way, the MGS follows a sampling distribution given by:
{\begin{equation}\label{eq:MGS_distribution}
	p(\hat{s}_{1},\dots,\hat{s}_{2K}|\mathbf{y}, \mathbf{H}) \sim \left(1-q\right) \psi \left(\alpha_1\right) + q \psi \left(\alpha_2\right)
	\end{equation}}
and
{\begin{equation}\label{eq:MGS_psi}
\hspace{0.14\linewidth}\psi \left(\alpha\right) = \exp \left(- \frac{||\mathbf{y} - \mathbf{H} \hat{\mathbf{s}} ||^2}{\alpha^2 \sigma^2}\right),
\end{equation}}
where $q$ denotes the mixing ratio. The MGS detector of \cite{DattaSept.2013} considers the $\alpha_1=1$, $\alpha_2=\infty$ combination, which results in a near-ML performance, overcoming the stalling problem of the GS, {being} also a simple implementation choice. {On the other hand, in high-order modulation, such as $64$-QAM and $256$-QAM, the noisy solution interferes in the algorithm's convergence, since there are a large number of symbols in the constellation and a simple random solution in this signal space has a high possibility of being far from the real solution, which causes the algorithm to require more iterations for convergence. In this sense, the proposed $d$-sMGS detector acts to mitigate this harmful effect.}

{Regarding the mixing ratio parameter $q$}, in \cite{DattaSept.2013} an analysis in low order QAM constellations is carried out and its suitable value choice is presented as the inverse of the number of dimensions in the system, {i.e.}, $q=\frac{1}{2K}$, {which is also employed in the proposed detector during our numerical simulations.}

In the MGS algorithm, an initial solution $\hat{\mathbf{s}}^{(t=0)}$ is {considered} for the estimated symbols vector, where $t$ represents the current iteration. {Indeed,} the initial solution may be {chosen either} by a random symbols vector or {as} the output of a linear low-complexity detector, such as zero forcing (ZF) or MMSE. The index $i$, in addition to the position of the vector $\hat{\mathbf{s}}$, also denotes the coordinate referring to the MGS algorithm, where $i=1,2,\dots,2K$. Therefore, each iteration requires $2K$ coordinate {updating}. At each iteration, updating the $2K$ coordinates is performed by sampling the distributions given by:
\begin{equation}\label{eq:conv_gibbs_update}
\hat{s}_{i}^{(t)} \sim p(\hat{s}_{i}|\hat{s}_1^{(t)},\dots,\hat{s}_{i-1}^{(t)},\hat{s}_{i+1}^{(t-1)},\dots,\hat{s}_{2K}^{(t-1)},\mathbf{y},\mathbf{H}).
\end{equation}
One can notice that by \eqref{eq:conv_gibbs_update} each updated coordinate is fed, in the same iteration, to the next coordinate. 

The probability of the $i$-th symbol assuming the value $a_j \in \mathbb{A}$, $\forall j = 1,\dots,|\mathbb{A}|$ can be written as:
\begin{equation}\label{eq:mgs_rew}
p(\hat{s}_{i}=a_j|\hat{\mathbf{s}}_{i-1}, \mathbf{y},\mathbf{H}) = \dfrac{\exp\left({- \dfrac{||\mathbf{y} - \mathbf{H} \mathbf{\hat{s}}_{i,j} ||^2}{\alpha^2\sigma^2}}\right)}{\sum_{l=1}^{|\mathbb{A}|} \exp\left({- \dfrac{||\mathbf{y} - \mathbf{H} \mathbf{\hat{s}}_{i,l} ||^2}{\alpha^2\sigma^2}}\right)},
\end{equation}
where the cardinality of set $\mathbb{A}$ is expressed as $|\mathbb{A}|$, while $\mathbf{\hat{s}}_{i,j}$ denotes the vector $\mathbf{\hat{s}}^{(t)}$ with its $i$-th position changed to the symbol $a_j$.

The sampling process based on \eqref{eq:mgs_rew} can lead to a numerical limitation due to the exponential function. In this sense, such implementation was carried out through a logarithmic intermediate step, as:
{\begin{align}\label{eq:mgs_practical_calc1}
& \log \left(p(\hat{s}_{i}=a_j|\hat{\mathbf{s}}_{i-1}, \mathbf{y},\mathbf{H})\right)  = \notag\\
&= \scriptstyle {f(i,j) - \left[f^{\mathrm{ord}}_0 + \log \left(1 + \sum_{m=1}^{|\mathbb{A}|-1} \exp \left( f^{\mathrm{ord}}_m - f^{\mathrm{ord}}_0 \right)\right)\right]} \notag \\
& = g(i,j) 
\end{align}}
where $f(i,j) = {- \dfrac{||\mathbf{y} - \mathbf{H} \mathbf{\hat{s}}_{i,j} ||^2}{\alpha^2\sigma^2}}$ and $f^{\mathrm{ord}}_i$ is $i$-th position of $\mathbf{f}$ in descending order, for $i=1,\dots,|\mathbb{A}|$. A practical and computationally efficient evaluation of MGS target Function is summarized in the Algorithm \ref{alg:mgs_calc}.

\begin{algorithm}[!htbp]
\caption{MGS Target Distribution Function Calculation} \label{alg:mgs_calc}
\small
\begin{algorithmic}[1]
	\STATE //\textit{Coordinate update process}
	\FOR{$i=1$ \textbf{to} $2K$}
		\STATE //\textit{{MGS target distribution function calculation}}
		\FOR{$j=1$ \textbf{to} $|\mathbb{A}|$}
		\STATE $f_{j} = \frac{||\mathbf{y} - \mathbf{H} \mathbf{\hat{s}}^{(t)}_{i,j} ||^2}{\alpha^2 \sigma^2}$
		\ENDFOR
		\STATE Ordinate $\mathbf{f}$ in descending order and denote $\mathbf{f}^{\mathrm{ord}}$
		\STATE $f' = {f}^{\mathrm{ord}}_1 + \log \left(1+ \sum_{m=2}^{|\mathbb{A}|} \exp \left( {f}^{\mathrm{ord}}_m - {f}^{\mathrm{ord}}_1\right)\right)$
		\FOR{$j=1$ \textbf{to} $|\mathbb{A}|$}
		\STATE $g_j = f_j - f'$
		\STATE $p(\hat{s}_{i}=a_j|\hat{\mathbf{s}}_{i-1}, \mathbf{y},\mathbf{H}) = \exp\left(g_j\right)$
		\ENDFOR
		\ENDFOR
		\STATE //\textit{Terminate}
	\end{algorithmic}
\end{algorithm}

The MGS algorithm ends after a certain amount of iterations, and the vector of estimated symbols is chosen as the vector that presented the lowest ML cost, considering all iterations. {In the next subsections, the additional strategy of multiple restarts (MR) \cite{DattaSept.2013} and the stopping criteria for the iterations and the restarts are addressed.}

\subsection{Multiple restarts}\label{subsec:mr}
In medium QAM order modulations, such as 16-QAM, the mixing strategy of MGS is unable to achieve near-optimal performance \cite{Damen2003} in a reasonable number of iterations, while MR procedure, as proposed in \cite{DattaSept.2013} has demonstrated promising results, leading the MGS-MR under 16-QAM to near-optimal performance. 

In {the aMGS and $d$-sMGS detectors the MR strategy is also incorporated, namely aMGS-MR and $d$-sMGS-MR detectors.} Thus, the {Algorithms \ref{alg:asmgs} and \ref{alg:d-smgs} run} either a maximal number of restarts $R_{\mathrm{max}}$ times or {it is} limited by a stopping criterion and the lowest cost found considering all restarts is the final solution. As discussed in Section \ref{sec:results}, the MR strategy can improve the convergence of the algorithm compared to the same number of iterations in a single execution, resulting in a better performance-complexity tradeoff.

\subsection{Stopping criterion}\label{subsec:stop_crit}
Given that the mixing strategy provides the local minimum escaping feature, the evolution of the cost function values across iterations becomes unpredictable and the optimal solution can be found before the maximum number of iterations $\mathcal{I}$ has been reached \cite{Mussi2018}. In this sense, an {\it efficient stopping criterion} is paramount in reducing the complexity of the MGS detector.

Similarly, the decision to set a restart in the algorithm requires a criterion definition, since the optimal solution may already have been found, not requiring an extra execution of the algorithm. Hence, MR strategy must be balanced aiming to achieve a better performance-complexity tradeoff.

Stopping criteria have been proposed in the literature. For instance, in \cite{DattaSept.2013}, the stopping criterion is based on the difference between the best ML cost found so far and the noise variance. Moreover, the QAM constellation size could be taken into account. The main idea in \cite{DattaSept.2013} is to stop the detection iterations if a maximum number of iterations $\mathcal{I}$ is attained or if the iteration in stalling mode is larger than a maximum of $\Theta_s$ iterations. 

Assume the estimated symbol vector, in the $t$-th iteration, is $\mathbf{\hat{s}}^{(t)}$. The {\it quality metric} of $\mathbf{\hat{s}}^{(t)}$ is defined as 
{\begin{equation}\label{eq:phi_quality_metric}
\hspace{.13\linewidth}\phi(\mathbf{\hat{s}}^{(t)}) = \dfrac{||\mathbf{y} - \mathbf{H} \mathbf{\hat{s}}^{(t)}||^2  -  N\sigma^2}{\sqrt{N}\sigma^2}.
\end{equation}}
Hence, the stalling limit for iterations, $\Theta_s$, is given by 
{\begin{equation}\label{eq:Theta_stalling_1}
\hspace{.18\linewidth}\Theta_s(\phi(\mathbf{\hat{s}}^{(t)})) = c_{s} \cdot  e^{\phi(\mathbf{\hat{s}}^{(t)})},
\end{equation}}
where $c_{s}$ is a constant depending upon the $M$-QAM constellation size, which increases with $M$. Although \eqref{eq:Theta_stalling_1} is suitable as a stopping criterion, a minimum number of iterations $c_{\mathrm{min}}$ {must be defined} to ensure the quality of symbol detection. Therefore, $\Theta_s$ can be rewritten as 
\begin{eqnarray}\label{eq:Theta_stalling_2}
\Theta_s(\phi(\mathbf{\hat{s}}^{(t)})) &=& \left\lceil \max \left(c_{\mathrm{min}},\,\, c_{s} \cdot e^{\phi(\mathbf{\hat{s}}^{(t)})}\right) \right\rceil, \nonumber \\
\text{with } \qquad c_{s} &=& c_1 \log_2 (M)  ,
\end{eqnarray}
where $c_1$ is a tunning constant which defines the allowed number of iterations in stalling mode.

For the MR strategy, the criterion set the allowable number of restarts $\Theta_r$, which also is based on quality metric $\phi(\mathbf{\hat{s}}^{(t)})$:
\begin{eqnarray}\label{eq:Theta_r}
\Theta_r(\phi(\mathbf{\hat{s}}^{(t)})) &=& \left\lceil \max \left(0,\,\, c_r \cdot \phi(\mathbf{\hat{s}}^{(t)}) \right) \right\rceil + 1, \nonumber\\
\text{whith}\quad c_r &=& c_2 \log_2 (M) \ ,
\end{eqnarray}
and $c_2$ is the tuning constant adjusting the maximum number of restarts.

At the end of each restart, $\Theta_r$ is computed and checked if the actual number of repetitions is less than $\Theta_r$. If yes, go to another run of the algorithm; else, output the solution vector with the minimum cost so far as the final solution. 

{For the aMGS and $d$-sMGS detectors presented below, aMGS and $d$-sMGS, we also assume the stop criteria described in this subsection.} 

\section{{Reducing the Impact of Noisy Solution}}\label{sec:noisy_appr}
{Originally, the mixture between the target distribution function solution and the random solution, proposed by MGS detector of \cite{DattaSept.2013}, attempted to escape local minima that degrade system performance. In fact, this procedure showed to significantly improves the performance, specially in low-order modulation scenarios, as $4$ or $16$-QAM. On the other hand, in high-order modulation systems, the large number of symbols causes the random solution to degrade the convergence of the algorithm since it is based on a coordinate update process which requires the global solution, thus one or more positions that consider a random solution (probably erroneous and far from the real solution) interfere in the convergence in the other positions and, consequently, in the global one. This condition is aggravated in high-dimension problems, {i.e.}, combining high-order modulations and number of antennas, which is the case of interest in this work. }

\begin{figure*}[!htbp]
	\centering
	\includegraphics[width=.95\textwidth]{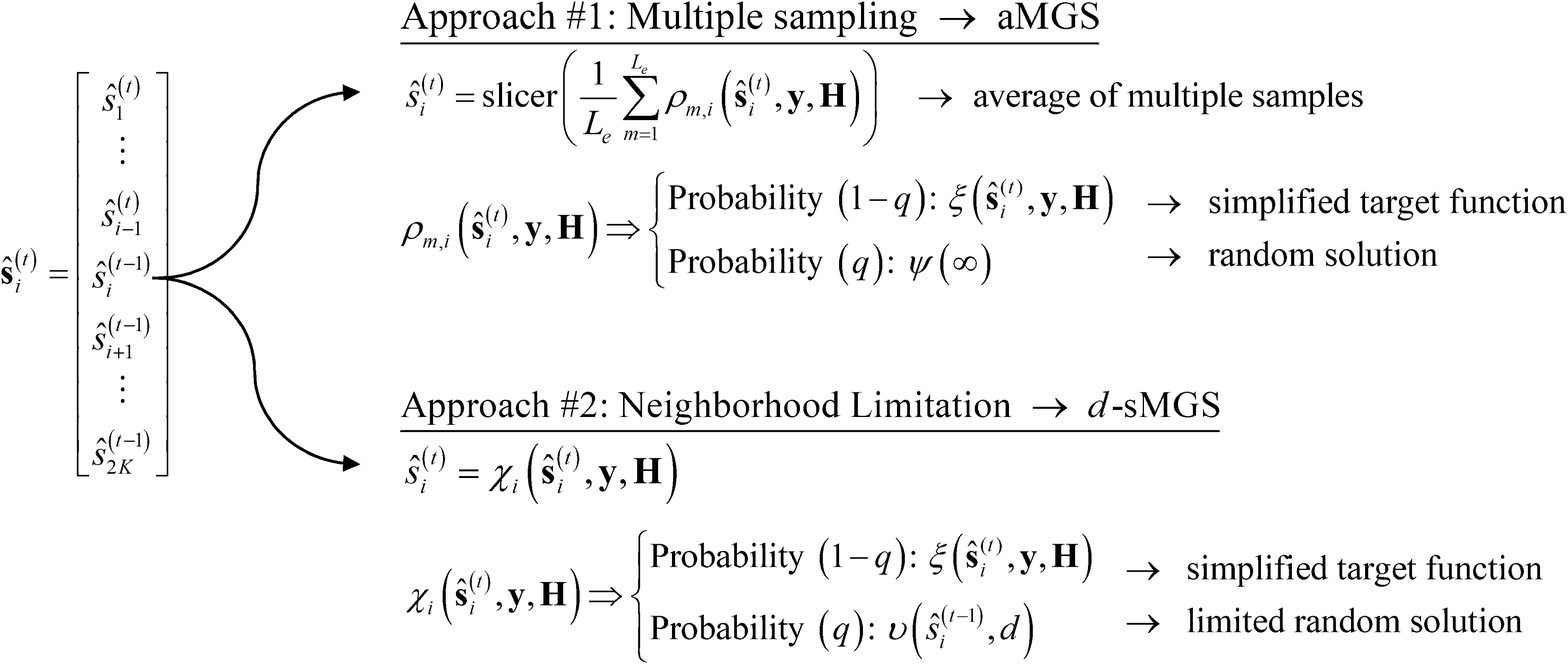}
	\caption{{A brief description of the coordinate update process on the aMGS and the proposed $d$-sMGS detectors.}}
	\label{fig:noisy_approaches}
\end{figure*}

{In this sense, two approaches that tries to alleviate the harmful impact of the noisy solution are described below. Fig. \ref{fig:noisy_approaches} summarizes the coordinate update process on the aMGS and $d$-sMGS detectors. {The strategy of multiple samples in mitigating the noisy solution also runs the risk of nullifying this solution if many samples are employed, this can happen since a mean among many terms from a r.v. with probabilities $q$ and $(1-q)$ -- with $q << (1-q)$ -- tends to be an average value in which the term with probability $q$ is nullified. In this sense, the noisy solution would be ineffective and the condition of stalling problem could happen, since the mixing of the MGS is a strategy to specifically tackle it.}}

\subsection{{Approach \#1:} Averaged MGS LS-MIMO detector}\label{subsec:amgs}
{The aMGS proposed in \cite{Mussi2018} is addressed herein and is based on the following improvements}:\\

\begin{enumerate}
	\item \textit{Averaged Multiple Sampling on each coordinate}: {differently from the single sampling strategy \cite{DattaSept.2013}, the aMGS employs an average between $L_e$ number of samples at each coordinate during the update process. By employing an averaged calculation, an intermediate (averaged) point between the target function symbol and the random symbol is more likely to be chosen, instead of a pure random symbol. As a result, the benefit of local minima escape is maintained, {whereas} the negative impact on the algorithm's convergence is smoothed.}\\
	
	\item \textit{Target Function Simplification}: to reduce the computational complexity related to target function calculation of \eqref{eq:mgs_rew}, the aMGS adopts a minimum ML cost approach. This simplification performs less mathematical operations, since the $||\mathbf{y} - \mathbf{H} \mathbf{\hat{s}}||$ computation is already performed in \eqref{eq:mgs_rew}. Thus, the aMGS target function, in the $t$-th iteration is evaluated as:
	\begin{eqnarray}\label{eq:smgs_target}
	\xi ({\mathbf{\hat{s}}_i^{(t)}}, \mathbf{y}, \mathbf{H} ) = \argmin_{j \in \left\{1, \dots, |\mathbb{A}|\right\}} ||\mathbf{y} - \mathbf{H} \mathbf{\hat{s}}_{i,j} || \ ,
	\end{eqnarray}
	where ${\mathbf{\hat{s}}_i^{(t)}}$ denotes the updated estimated symbol vector until the $(i-1)$ position {at the $t$-th iteration}, whereas the other remaining  $i, \, (i+1), \dots,\,  2K$ positions assume the values from the previous iteration, {i.e.}, 
	$$
	{\mathbf{\hat{s}}_i^{(t)}} = [\hat{s}_1^{(t)},\dots,\hat{s}_{i-1}^{(t)},\hat{s}_{i}^{(t-1)},\dots,\hat{s}_{2K}^{(t-1)}]^T
	$$
\end{enumerate}

	Compared to \eqref{eq:mgs_rew}, the calculation of \eqref{eq:smgs_target} performs less operations while achieves same BER performance \cite{Mussi2018}.
	
\subsubsection{{MS in} coordinate update process}
The coordinate update process of aMGS is defined by:
{\begin{equation}\label{eq:aMGS_update}
\hspace{.14\linewidth}\hat{s}_{i}^{(t)} = \frac{1}{L_{\mathrm{e}}} \sum_{m=1}^{L_{\mathrm{e}}} {\rho_{m,i} (\hat{\mathbf{s}}^{(t)}_{i},\mathbf{y},\mathbf{H})} \ ,
\end{equation}}
where $L_e$ is the number of samples (realizations), and the random variable (r.v.) $\rho_{m,i}$ is a mixture of two r.v. with weight given by the mixing ratio $q$, defined by: 
\begin{equation}\label{eq:aMGS_distribution}
{\rho_{m,i} (\hat{\mathbf{s}}^{(t)}_{i}, \mathbf{y}, \mathbf{H})} \sim
\left(1-q\right) \cdot \xi ({\mathbf{\hat{s}}_i^{(t)}}, \mathbf{y}, \mathbf{H})+ q \cdot \psi \left(\infty \right) .
\end{equation}
It is important to note that, being \eqref{eq:smgs_target} a deterministic function, during the $L_e$ realizations on each coordinate, \eqref{eq:smgs_target} is calculated only once, when $m=1$. After that, each $m$ realization has the computational cost of generating a random number (relative to the mixing ratio). 

At the end of algorithm iterations, the vector with the lowest cost is assumed the best global solution. Due to the mean operation,  a slicer for $M$-QAM constellation is needed at the end of the detection procedure. Thus,
{\begin{equation}
\mathbf{\hat{s}}_{\mathrm{best}} = {\mathrm{slicer}}(\mathbf{\hat{s}}_{\mathrm{f-best}} ),
\end{equation}}
where $\mathbf{\hat{s}}_{\mathrm{f-best}} $ is the ``floating-best" solution which represents the estimated vector related to the best global cost {attained after $\mathcal{I}$ iterations,} and  $\mathbf{\hat{s}}_{\mathrm{best}} $ is the final estimated symbol vector.  A pseudocode for the aMGS is described in Algorithm \ref{alg:asmgs}. 
\begin{algorithm}[!htbp]
\caption{{aMGS  for {LS-}MIMO detection}} \label{alg:asmgs}
\small
\singlespace
\begin{algorithmic}[1]
\STATE{Initialization}  $\mathbf{s}^{(t=0)}:$ initial random vector; $L_e$ \# samples; $\mathcal{I}$: max. number of iterations; $t=1$; $q$: mixing ratio; $\mathbb{A} = \left\{ a_1, a_2, \dots, a_{|\mathbb{A}|} \right\}$
		\STATE \textit{{//}{Iterative process}}
		\WHILE{$t < \mathcal{I}$} 
		\STATE \textit{{//}Coordinate update process}
		\FOR{$i=1$ \textbf{to} $2K$}
		\STATE \textit{{//}{Simplified target function calculation}}
		\FOR{$j=1$ \textbf{to} $|\mathbb{A}|$}
		\STATE $f_{j} = ||\mathbf{y} - \mathbf{H} \mathbf{\hat{s}}^{(t)}_{i,j} ||$
		\ENDFOR
		\STATE $f_{\mathrm{min}} = \argmin_{j} f_j$
		\STATE ${\xi ({\mathbf{\hat{s}}_i^{(t)}}, \mathbf{y}, \mathbf{H})} = a_{{f}_{\mathrm{min}}}$ 
		\STATE {//} $L_e$ \textit{{samples on each coordinate}}
		\FOR{${m}=1$ \textbf{to} $L_e$}
		\STATE generate $u_{i,{m}} \sim U[0,1]$
		\IF{($u_{i,{m}} > q$)}
		\STATE ${\rho_{m,i} (\hat{\mathbf{s}}^{(t)}_{i}, \mathbf{y}, \mathbf{H})} = \xi ({\mathbf{\hat{s}}_i^{(t)}}, \mathbf{y}, \mathbf{H})$
		\ELSE
		\STATE $r \sim {\lfloor}(U[1,|\mathbb{A}|]){\rceil}$
		\STATE ${\rho_{m,i} (\hat{\mathbf{s}}^{(t)}_{i}, \mathbf{y}, \mathbf{H})} = a_r$
		\ENDIF
		\ENDFOR
		\STATE \textit{{//}{Averaging between samples}}
		\STATE $\hat{s}_i^{(t)} = \frac{1}{L_e} \sum_{{m}=1}^{L_e} {{\rho_{m,i} (\hat{\mathbf{s}}^{(t)}_{i}, \mathbf{y}, \mathbf{H})}}$
		\STATE \textit{{//}{Storage of cost and temporary vectors}}
		\STATE $\beta_i = ||\mathbf{y} - \mathbf{H} \mathbf{\hat{s}}^{(t)}_{i} ||$
		\STATE $\mathbf{S}_{:,i} = \mathbf{\hat{s}}^{(t)}_{i}$
		\ENDFOR
		\STATE \textit{{//}{Best cost in the $t$-th iteration}}
		\STATE $\beta_{\mathrm{min}} = \min \beta_i$
		\STATE \textit{{//}{Best global solution test}}
		\IF{$(\beta_{\mathrm{min}} < \beta_{\mathrm{best}})$}
		\STATE $\beta_{\mathrm{best}} = \beta_{\mathrm{min}}$
		\STATE $i_{\mathrm{min}} = \argmin_i \beta_i$
		\STATE $\mathbf{\hat{s}}_{\mathrm{f-best}}  = \mathbf{S}_{:,i_{\mathrm{min}} }$
		\ENDIF
		\STATE $t = t + 1$
		\STATE ${b}_t = \beta_{\mathrm{best}}$
		\STATE \textit{{//}{Stop criterion for iterations}}
		\IF{$(b_t == b_{t-1})$}
		\STATE $m = \Theta_s (\mathbf{\hat{s}}_{\mathrm{best}} )$
		\IF{$(m < t)$}
		\IF{$(b_t == b_{t-m})$}
		\STATE $\mathbf{\hat{s}}_{\mathrm{best}}  = {\mathrm{slicer}}(\mathbf{\hat{s}}_{\mathrm{f-best}}  )$
		\STATE \textit{{//}Terminate}
		\ENDIF
		\ENDIF	
		\ENDIF		
		\ENDWHILE
\STATE $\mathbf{\hat{s}}_{\mathrm{best}}  = {\mathrm{slicer}}(\mathbf{\hat{s}}_{\mathrm{f-best}})$
\STATE \textit{{//}Terminate}
\end{algorithmic}
\end{algorithm}

\subsection{Approach \#2: Simplified MGS with Neighborhood Limitation LS-MIMO detector}\label{sec:d-smgs}
{We propose an different approach which is based on a neighborhood limitation of distance $d$ in the random solution and is named $d$-sMGS LS-MIMO detector. The term simplified refers to the simplified target function of Eq. \ref{eq:smgs_target}, which is also employed in this scheme. }

{The proposed $d$-sMGS detector acts in the symbol constellation performing a NL, with distance $d$ in relation to the symbol estimated in the previous iteration, when sorting the random symbol. This procedure showed to significantly improves the convergence when a modulation of high-order is considered, as disposed in section \ref{sec:results}, and presents the lowest per-symbol complexity among MGS and aMGS, since it considers the simplified target function (overcoming the MGS in mathematical operations) and performs a single sample (overcoming the multiple sampling aMGS), as showed in section \ref{sec:complexity}.}

\subsubsection{{NL in coordinate update process}}
{The $d$-sMGS coordinate update process is based on a mixture between the simplified target function, Eq. \ref{eq:smgs_target}, and a limited random solution. Thus, the estimated symbol in the $t$-iteration at the $i$-th coordinate is given by:}
{\begin{equation}\label{eq:d-sMGS_update}
\hspace{.23\linewidth}\hat{s}_{i}^{(t)} = {\chi_{i} (\hat{\mathbf{s}}^{(t)}_{i},\mathbf{y},\mathbf{H})} \ ,
\end{equation}}
{where $\chi_{i}\left(\cdot\right)$ is the mixed r.v. with weight $q$, defined by:}
{\begin{equation}\label{eq:d-sMGS_distribution}
\chi_{i} (\hat{\mathbf{s}}^{(t)}_{i}, \mathbf{y}, \mathbf{H}) \sim
\left(1-q\right) \cdot \xi ({\mathbf{\hat{s}}_i^{(t)}}, \mathbf{y}, \mathbf{H})+ q \cdot \upsilon \left( \hat{s}_i^{(t-1)}, d\right) \ ,
\end{equation}}
{the r.v. $\upsilon \left( \hat{s}_i^{(t-1)}, d\right)$ denotes an uniform sorted symbol in the constellation neighborhood of $\hat{s}_i^{(t-1)}$, with distance $d$.} 

{In this algorithm, the neighborhood of the current solution $\hat{s}_i^{(t-1)}$ is defined as}
{\begin{equation}\label{eq:neigh_distance}
\mathcal{N} \left( \hat{s}_i^{(t-1)}, d\right) = \left\{ {s}' \in \mathbb{A} \ | \ \kappa_d \left(\hat{s}_i^{(t-1)},{s}'\right) \leq d \right\},
\end{equation}}
{where $\kappa_d$ is the symbol distance function in the real-valued constellation considered, for example, let $\mathbb{A}=\left\{ -7,-5,-3,-1,+1,+3,+5,+7 \right\}$, $\hat{s}_i^{(t-1)}=-3$ and $s'=+1$, then the symbol distance function results in $\kappa_d \left(\hat{s}_i^{(t-1)},{s}'\right)=2$. }

{Thus, the r.v. $\upsilon \left( \hat{s}_i^{(t-1)}, d\right)$ samples from a discrete uniform distribution on the set $\mathcal{N} \left( \hat{s}_i^{(t-1)}, d\right) = \left\{  n_1, \dots, n_{|\mathcal{N}|} \right\}$.}

{A pseudocode for the proposed $d$-sMGS is described in Algorithm \ref{alg:d-smgs}. The multiple restarts additional strategy is omitted, since it simply restarts the algorithm with another initial solution.}

\begin{algorithm}[!htbp]
	\caption{{$d$-sMGS  for {LS-}MIMO detection}} \label{alg:d-smgs}
	\small
	{
	\begin{algorithmic}[1]
		\STATE \textit{{//}{Initialization}}
		\STATE $\mathbf{s}^{(t=0)}:$ initial random vector; $d$: constellation distance; $\mathcal{I}$: max. number of iterations; $t=1$; $q$: mixing ratio; $\mathbb{A} = \left\{ a_1, a_2, \dots, a_{|\mathbb{A}|} \right\}$
		\STATE \textit{{//}{Iterative process}}
		\WHILE{$t < \mathcal{I}$}
			\STATE \textit{{//}Coordinate update process}
			\FOR{$i=1$ \textbf{to} $2K$}
				\STATE //\textit{Evaluation of $\chi_{i}\left(\cdot\right)$, Eq. \ref{eq:d-sMGS_distribution} }
				\STATE generate $u_{i} \sim U[0,1]$
				\IF{($u_{i} > q$)}
					\STATE \textit{{//}{Simplified target function calculation, Eq. \ref{eq:smgs_target}}}
					\FOR{$j=1$ \textbf{to} $|\mathbb{A}|$}
						\STATE $f_{j} = ||\mathbf{y} - \mathbf{H} \mathbf{\hat{s}}^{(t)}_{i,j} ||$
					\ENDFOR
					\STATE $f_{\mathrm{min}} = \argmin_{j} f_j$
					\STATE $\xi ({\mathbf{\hat{s}}_i^{(t)}}, \mathbf{y}, \mathbf{H}) = a_{{f}_{\mathrm{min}}}$ 
					\STATE $\chi_{i} (\hat{\mathbf{s}}^{(t)}_{i}, \mathbf{y}, \mathbf{H}) = \xi ({\mathbf{\hat{s}}_i^{(t)}}, \mathbf{y}, \mathbf{H})$
				\ELSE
					\STATE //\textit{Generation of the $d$-limited set}
					\STATE $\mathcal{N} \left( \hat{s}_i^{(t-1)}, d\right) = \left\{ {s}' \in \mathbb{A} \ | \ \kappa_d \left(\hat{s}_i^{(t-1)},{s}'\right) \leq d \right\}$
					\STATE //\textit{Sampling from a discrete uniform distribution on the set $\mathcal{N} \left( \hat{s}_i^{(t-1)}, d\right)=\left\{n_1,\dots,n_{|\mathcal{N}|}\right\}$}
					\STATE $\upsilon \left( \hat{s}_i^{(t-1)}, d\right) \sim \mathcal{U} \left[n_1,n_{|\mathcal{N}|}\right]$
					\STATE $\chi_{i} (\hat{\mathbf{s}}^{(t)}_{i}, \mathbf{y}, \mathbf{H}) = \upsilon \left( \hat{s}_i^{(t-1)}, d\right)$
				\ENDIF
				\STATE //\textit{Updating the estimated symbol vector in the $i$-position}
				\STATE $\hat{s}_i^{(t)} = \chi_{i} (\hat{\mathbf{s}}^{(t)}_{i}, \mathbf{y}, \mathbf{H})$
			\ENDFOR
			\STATE \textit{{//}{Storage of cost and temporary vectors}}
			\STATE $\beta_i = ||\mathbf{y} - \mathbf{H} \mathbf{\hat{s}}^{(t)}_{i} ||$
			\STATE $\mathbf{S}_{:,i} = \mathbf{\hat{s}}^{(t)}_{i}$
			\STATE \textit{{//}{Best cost in the $t$-th iteration}}
			\STATE $\beta_{\mathrm{min}} = \min \beta_i$
			\STATE \textit{{//}{Best global solution test}}
			\IF{$(\beta_{\mathrm{min}} < \beta_{\mathrm{best}})$}
				\STATE $\beta_{\mathrm{best}} = \beta_{\mathrm{min}}$
				\STATE $i_{\mathrm{min}} = \argmin_i \beta_i$
				\STATE $\mathbf{\hat{s}}_{\mathrm{best}}  = \mathbf{S}_{:,i_{\mathrm{min}} }$
			\ENDIF
			\STATE $t = t + 1$
			\STATE ${b}_t = \beta_{\mathrm{best}}$
			\STATE \textit{{//}{Stop criterion for iterations}}
			\IF{$(b_t == b_{t-1})$}
				\STATE $m = \Theta_s (\mathbf{\hat{s}}_{\mathrm{best}} )$
				\IF{$(m < t)$}
					\IF{$(b_t == b_{t-m})$}
						\STATE \textit{{//}Terminate}
					\ENDIF
				\ENDIF	
			\ENDIF		
		\ENDWHILE
		\STATE \textit{{//}Terminate}
	\end{algorithmic}}
\end{algorithm}

\section{Computational Complexity}\label{sec:complexity}
The computational complexity is described in terms of real number of operations {({\it rops})}, {in} which one {\it rop} denotes the computational complexity of the real mathematical operations: addition, subtraction, multiplication or division. For the exponential and logarithmic functions, an approximation through Taylor Series with 18 terms has been considered to calculate the computational complexity. {Table \ref{tab:complexity} describes the per-symbol computational complexity ($\mathcal{C}_T$) involved in each step of $d$-sMGS algorithm.} Additionally, the total per-symbol complexity of the aMGS and the conventional MGS has been evaluated. {The per-symbol complexity of the initial solution is denoted by $\mathcal{C}_I$, which is adopted in this work as the output of an MMSE detector, which has also its total complexity described in the Table \ref{tab:complexity} \cite{Liu2009}}. 
{From Table \ref{tab:complexity}, one can notice that the $d$-sMGS algorithm, aMGS and MGS algorithms have the same asymptotic per-symbol complexity order of $\mathcal{O}(K^2)$, although the conventional MGS algorithm may require an additional complexity dependent on constellation size due to the exponential function, which is represented by the cardinality $|\mathbb{A}|$. On the other hand, the additional complexity due to the averaged strategy of the aMGS represents a negligible impact, since it requires only $(2L_e+2)$ {\it rops} per iteration, {{whereas} such additional complexity is} not dependent on the problem size. {The proposed $d$-sMGS algorithm combines advantages of both by using a single sample such as the MGS and the simplified aMGS target function. The complexity increment given by the neighborhood constraint is considered negligible, since the symbol is already previously estimated and such procedure represents only a random sampling in a restricted vector.}}

{\renewcommand{\arraystretch}{1.6}%
\begin{table*}[!htbp]
\caption{Per-symbol Computational Complexity of aMGS, conventional MGS and MMSE Algorithms.}
\small 
\begin{center}
			\begin{tabular}{rcc}
				\hline
				\textbf{Procedure} &  {\bf Step} & \textbf{Complexity}  \\ 
				\hline\hline 
				\multicolumn{3}{c}{{\bf $d$-MGS --  Algorithm  \ref{alg:d-smgs}}}\\ 
				\hline
				{Target function calculatio}n & {{lines 11--16}} & \multicolumn{1}{c}{{$16KN - 4N + |\mathbb{A}|\left(16N+2\right)$}} \\ 
				\multicolumn{1}{r}{{Generation of the $d$-limited set}}  &  {{line 19}}  & \multicolumn{1}{c}{{\textit{negligible}}} \\ 
				\multicolumn{1}{r}{{Cost computation at each coordinate}} &  {{line 28}} &\multicolumn{1}{c}{{$20N$}} \\ 
				\multicolumn{1}{r}{{$\Theta_s$,  \,\, {Eq. \eqref{eq:Theta_stalling_2}}} } &  {{line 41}}  & \multicolumn{1}{c}{{$\frac{24}{K}$}} \\ 
				\hline
				\multicolumn{1}{r}{{\textbf{Total per-symbol complexity}:}} & \multicolumn{2}{l}{{$\mathcal{C}_T = \mathcal{C}_I + {\mathcal{I}_{\mathrm{eff}}}\left[16KN+16N+|\mathbb{A}|\left(16N+2\right) + \frac{24}{K}\right]$}} \\
				\hline
				\\[-2mm]
				\hline
				\multicolumn{3}{c}{\bf aMGS --  Algorithm  \ref{alg:asmgs}}\\ 
				\hline
				Target function calculation & {lines 8--12} & \multicolumn{1}{c}{$16KN - 4N + |\mathbb{A}|\left(16N+2\right)$} \\ 
				\multicolumn{1}{r}{Averaging between samples}  &  {line 24}  & \multicolumn{1}{c}{$2L_e + 2$} \\ 
				\multicolumn{1}{r}{Cost computation at each coordinate} &  {line 26} &\multicolumn{1}{c}{$20N$} \\ 
				\multicolumn{1}{r}{$\Theta_s$,  \,\, {Eq. \eqref{eq:Theta_stalling_2}} } &  {line 41}  & \multicolumn{1}{c}{$\frac{24}{K}$} \\ 
				\hline
				\multicolumn{1}{r}{\textbf{Total per-symbol complexity}:} & \multicolumn{2}{l}{$\mathcal{C}_T = \mathcal{C}_I + {\mathcal{I}_{\mathrm{eff}}}\left[16KN+16N+|\mathbb{A}|\left(16N+2\right)+(2L_e+2) + \frac{24}{K}\right]$} \\
				\hline
				\\[-2mm]
				\hline
				\multicolumn{3}{c}{{\bf MGS --  {Target distribution function calculation on Algorithm}  \ref{alg:mgs_calc}}}\\ 
				\hline
				{Target distribution function calculation} & {lines 4--6} & \multicolumn{1}{c}{{$16KN - 4N + |\mathbb{A}|\left(16N+12\right)$}} \\ 
				\multicolumn{1}{r}{{Evaluation of each symbol probability}}  &  {lines 8--12}  & \multicolumn{1}{c}{{$1238|\mathbb{A}|$}} \\ 
				\multicolumn{1}{r}{{Cost computation of estimated vector}} &  &\multicolumn{1}{c}{{$\frac{10N}{K}$}} \\ 
				\multicolumn{1}{r}{{$\Theta_s$,  \,\, {Eq. \eqref{eq:Theta_stalling_2}} }} &  & \multicolumn{1}{c}{{$\frac{24}{K}$}} \\ 
				\hline
				\multicolumn{1}{r}{\textbf{Total per-symbol complexity}:} & \multicolumn{2}{l}{{$\mathcal{C}_T = \mathcal{C}_I + {\mathcal{I}_{\mathrm{eff}}}\left[16KN-4N+|\mathbb{A}|\left(16N+1450\right) + \frac{10N+24}{K} \right]$}} \\
				\hline
				\\[-2mm]
				\hline
				\multicolumn{3}{c}{{\bf MMSE Algorithm}}\\ 			
				\hline
				\multicolumn{1}{r}{\textbf{Total per-symbol complexity}:} & \multicolumn{2}{l}{$\mathcal{C}_T = \left(\frac{1}{6}\right) K^2 + \left(\frac{3}{2}\right)NK + \left(\frac{3}{2}\right)N + \left(\frac{5}{6}\right) $}\\
				\hline
			\end{tabular}
		\end{center}
		\label{tab:complexity}
\end{table*}}

{From Table \ref{tab:complexity}, it may be noted that the proposed $d$-sMGS has its per-symbol complexity independent of the parameter $d$, so the use of larger neighborhoods in the random symbol generation has no impact on complexity. With respect to the per-symbol complexity of the initial solution, $\mathcal{C}_I$, in this work we adopted the output of an MMSE detector, which has also its total complexity described in the Table \ref{tab:complexity}.}

It is important to emphasise that the complexity of the {$d$-sMGS}, aMGS and MGS algorithms is defined by the number of iterations, which is controlled by the stopping criterion $\Theta_s$, with the upper limit $\mathcal{I}$. Similarly, the amount of restarts is controlled by $\Theta_r$, with an upper limit $R_{\rm max}$. In terms of complexity, the MR procedure can be interpreted as an extra amount of iterations necessary for each new restart. {In this sense, an ${\mathcal{I}_{\rm eff}}$ is considered in Table \ref{tab:complexity}, which denotes the total amount of iterations (including all restarts) performed at each symbol period. Since Monte-Carlo method is employed in simulations, in section \ref{sec:results} a mean value of ${\mathcal{I}_{\rm eff}}$ considering all realisations is evaluated and is called {\it effective number of iterations} (ENI):}
{\begin{equation}
	\hspace{.23\linewidth}{\rm ENI} = {\frac{1}{T}} \sum_{i=1}^{{T}} {\mathcal{I}_{{\rm eff},i}} \ ,
\end{equation}}
{where $T$ denotes the total {number of realisations (symbol periods) during the Monte-Carlo method simulation} and ${\mathcal{I}_{{\rm eff},i}}$ denotes the ${\mathcal{I}_{\rm eff}}$ in the $i$-realisation.}

\subsection{{Quality metric}}\label{subsubsec:tradeoff}
Due to the large number of parameters involved in the presented LS-MIMO detectors, a simple performance-complexity tradeoff metric is considered \cite{Mussi2018}, which aims to establish a fair comparison analysis between different detection strategies:

\begin{equation}
\chi(\mathrm{BER}, \mathcal{C}_T) = -\frac{10 \log_{10} \left({\mathrm{BER}}\right)}{10^{-8}\cdot \mathcal{C}_T}  = -\frac{\mathrm{BER}_{\mathrm{dB}}}{10^{-8}\cdot\mathcal{C}_T}
\end{equation}
where 
$\mathrm{BER}_{\mathrm{dB}}$ denotes the bit error rate in dB. Higher values of $\chi(\cdot)$ imply more efficient and effective LS-MIMO detector.

\section{Numerical results and discussion}\label{sec:results}
In this section the uncoded BER performance related to the $d$-sMGS algorithm for LS-MIMO detection is evaluated through Monte Carlo simulations. The simulations are performed for a large-scale MIMO operating in multiplexing mode and assuming that a perfect channel state information is available at the receiver side. Table \ref{tab:sys_par} summarizes the main system and channel parameter values deployed in this section. 

{As proposed in \cite{DattaSept.2013}, the mixing ratio parameter is adopted as the inverse of the number of dimensions in the system, {i.e.}, $q=\frac{1}{2K}$. For the stopping criterion parameters, we have adopted $c_1=10$,  $c_2=1.0$, and $c_{\min}=10$ \cite{Mussi2018}. }

This numerical simulation section has been divided into two main parts:
{in subsection \ref{subsec:amgs_analysis} the mixing ratio $q$ and number of samples $L_e$ parameters of the aMGS detector are discussed, as it denotes a technique that also aims at reducing the impact of the noisy solution; in subsection \ref{subsec:d_smgs_results}, we present numerical results of performance and computational complexity of the proposed $d$-sMGS detector against the aMGS and MGS techniques, addressed in this work.}

{\renewcommand{\arraystretch}{1.4}%
	\begin{table}[!htbp]
		\caption{LS-MIMO system and channel parameters.}%
\begin{center}
			\small
			\begin{tabular}{ll}
				\hline
				\bf Parameter & \bf Value\\
				\hline\hline
				\multicolumn{2}{c}{{\it LS-MIMO System}} \\
				\hline
				Link direction & Uplink (UL)\\
				\# Rx antennas (BS) & $N \in \{	64, 128\}$\\
				\# Tx antennas (MTs) & \multirow{2}{*}{$K \in \{48, 96\}$} \\
				(single user-antenna) & \\
				System loading & $\beta = \frac{K}{N}  \in [0.3125,0.90625]$\\
				{Modulation order} & 64-QAM \\
				{SNR ranges} & {$\gamma_{\mathrm{dB}} \in \left[0, \,\, 25\right]$ dB}\\
				\hline
				\multicolumn{2}{c}{{\it Channel}} \\
				\hline
				Channel type & Flat Rayleigh \\
				Channel availability & Perfectly known at receiver \\
				\hline
				\multicolumn{2}{c}{{\it Specific detector parameters}} \\
				\hline
				Max. number of iterations & $\mathcal{I} = 8 K \sqrt{M}$\\
				Max. number of restarts & {$R_{\mathrm{max}} = 20$}\\
				{NL distance} & {{$d \in \left\{1 , 2, 3\right\}$}}\\
				Mixing ratio & {$q =\frac{1}{2K}$} \\
				{Stop criterion parameters} & {$c_1=10$; $c_2=1$; $c_{\mathrm{min}} = 10$} \\
				\hline
			\end{tabular}
		\end{center}
		\label{tab:sys_par} 
	\end{table}}

\subsection{aMGS parameters discussion}\label{subsec:amgs_analysis}

\begin{figure*}[!htbp]
	\centering
	\begin{subfigure}[$96\times128$, $L_e=1$]{\includegraphics[width=.49\textwidth]{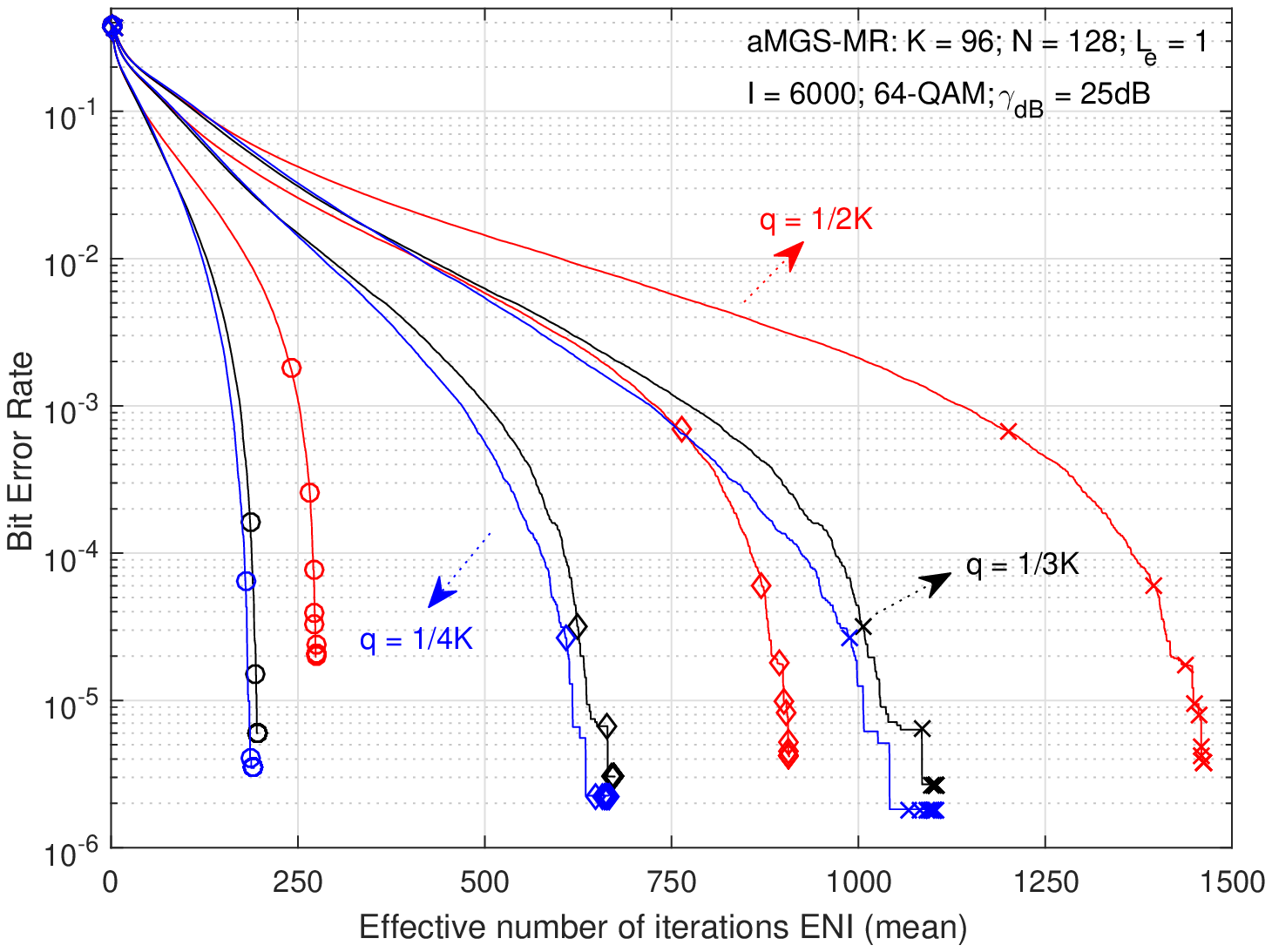}}
	\end{subfigure}
	\begin{subfigure}[$96\times128$, $L_e=2$]{\includegraphics[width=.49\textwidth]{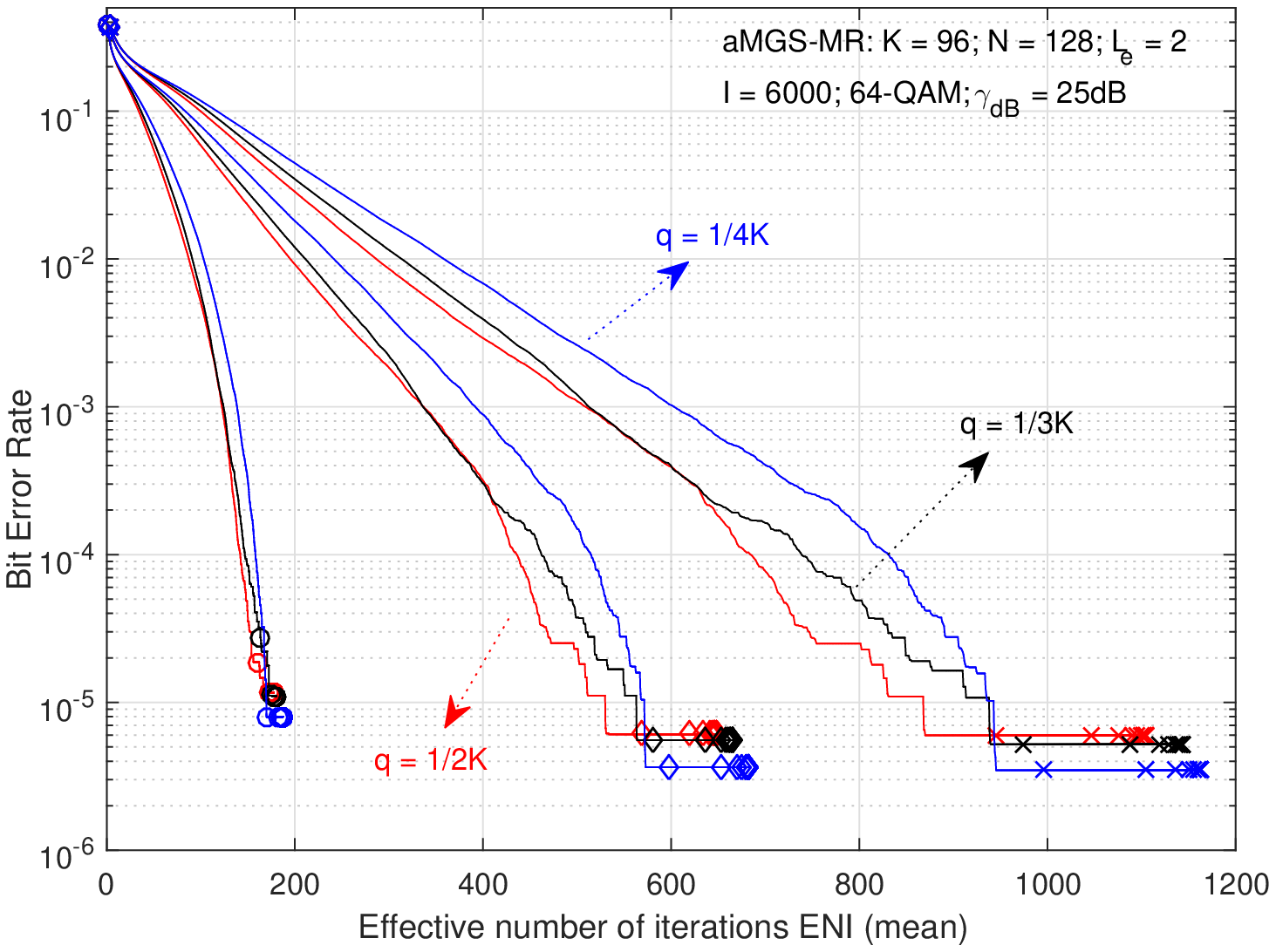}}
	\end{subfigure}
	\begin{subfigure}[$96\times128$, $L_e=4$]{\includegraphics[width=.49\textwidth]{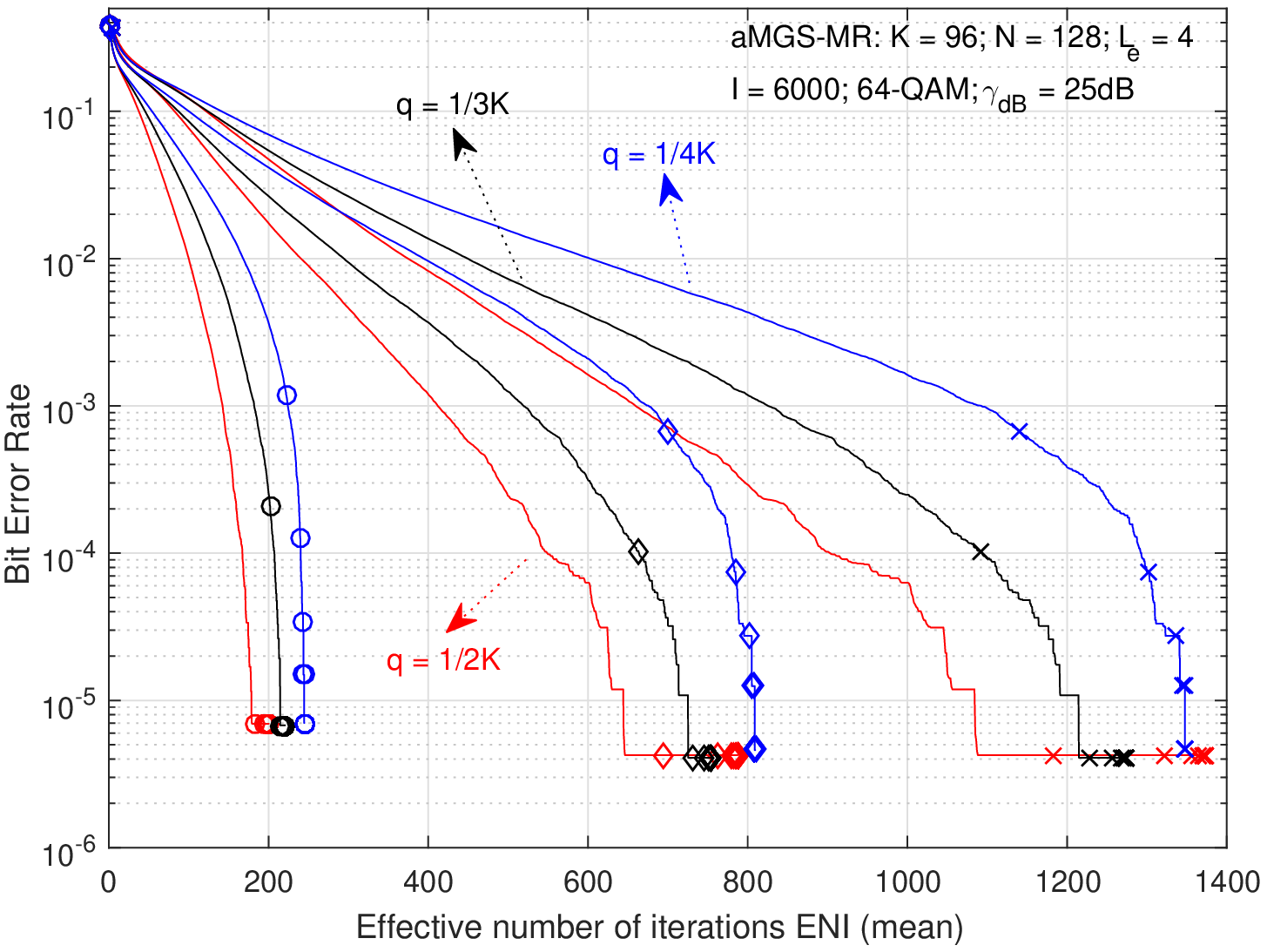}}
	\end{subfigure}
	\begin{subfigure}[$96\times128$, $L_e=8$]{\includegraphics[width=.49\textwidth]{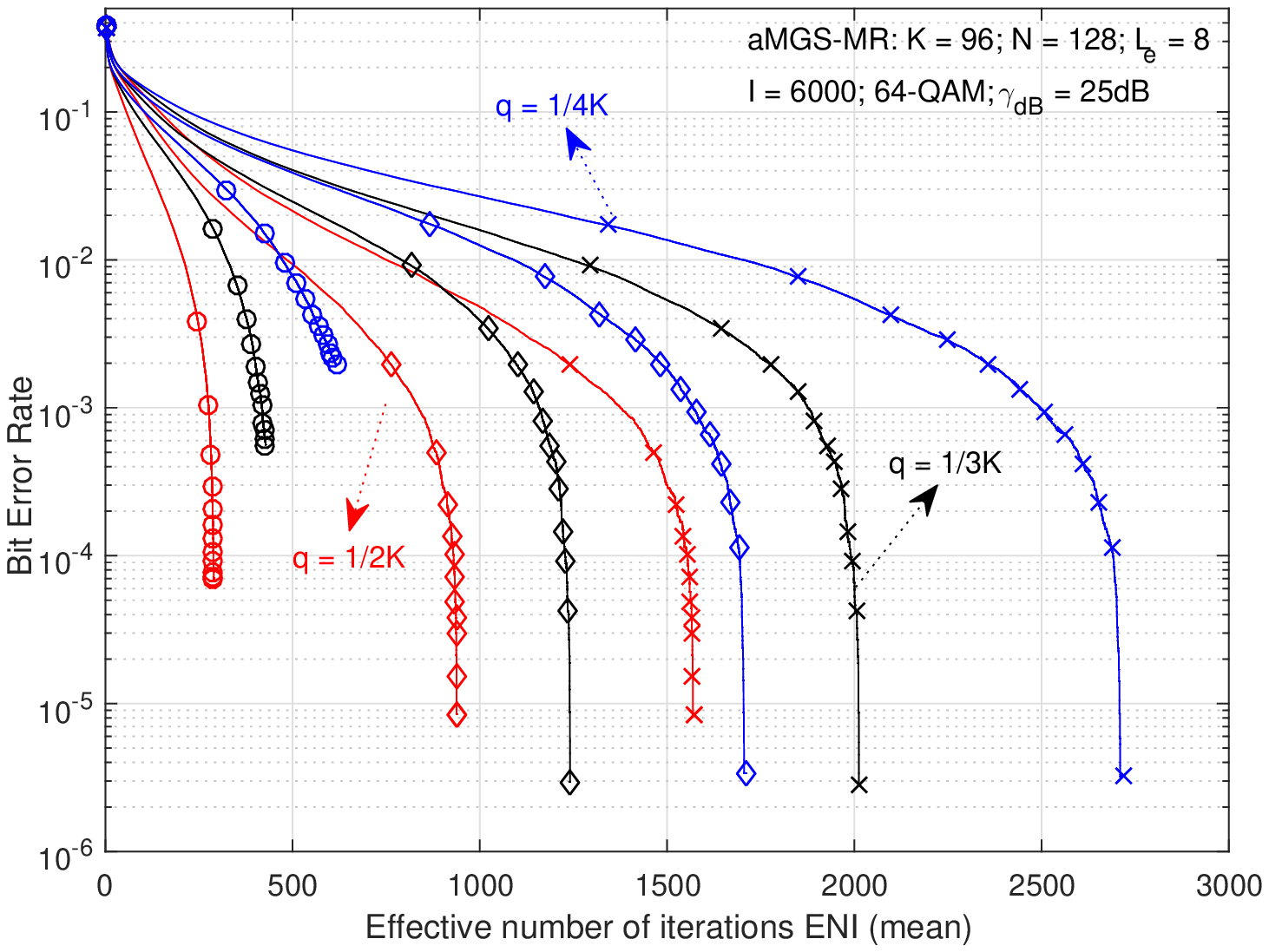}}
	\end{subfigure}
	{\includegraphics[width=.49\textwidth]{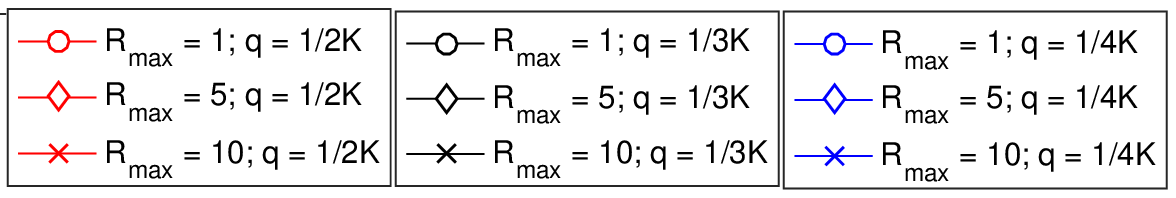}}
	\caption{BER performance convergence of different mixing ratios, $q$, of aMGS in medium number of antennas scenario ($K=96$, $N=128$) at $\gamma_{\mathrm{dB}}=25$dB, $64$-QAM, $R_{\mathrm{max}} = \{ 1, 5, 10\}$, $\mathcal{I} = 6000$ and different $L_e$ samples: {\bf a)} $L_e=1$, {\bf b)} $L_e=2$, {\bf c)} $L_e=4$ and {\bf d)} $L_e=8$. \cite{Mussi2018}}
	\label{fig:QJointed_convergence_96x128_64QAM_25dB}
\end{figure*}

{The aMGS-MR BER performance for different mixing ratios $q =  \{1/2K , 1/3K, 1/4K\}$, considering $R_{\mathrm{max}} = \{ 1, 5, 10\}$, is presented in Fig. \ref{fig:QJointed_convergence_96x128_64QAM_25dB} for each fixed $L_e\in \{1,\,  2,\, 4, \,  8\}$ samples scenario \cite{Mussi2018}. The number of users is equal to $K=96$ while $N=128$ BS antennas ($\beta=0.75$). The system is operating {under medium-high SNR,} $\gamma_{\mathrm{dB}}=25$dB.  
First, it is evident that the choice of different mixing ratio values impact both performance and complexity (represented by the ENI quantity at convergence). In addition, one can notice that the large amount of $L_e = 8$ samples becomes harmful to the algorithm, once convergence is achieved with larger ENI. Among the other results, the best performance-complexity tradeoff is presented with $L_e = 2$ samples and $q=1/4K$, which results in: $\left. \chi \right|_{L_e=2} = 44.89$; against $\left. \chi \right|_{L_e=4} = 37.85$ with 4 samples and $q=1/2K$; and $\left. \chi \right|_{L_e=1} = 39.66$ with 1 sample and $q=1/4K$. A detailed analysis of the aMGS performance/complexity gain in relation to the mixing ratio and the number of samples can be found in \cite{Mussi2018}.}

{It can also be concluded that with increasing number of samples $L_e$, the curve represented by $q = 1 / 2K$ has its convergence improved, resulting in less complexity. That is, when the impact of the noisy solution is reduced, the choice of $q = 1 / 2K$ is presented as the best performance-complexity tradeoff.} {In this sense, the value $q = 1 / 2K$ is adopted for the proposed detector $d$-sMGS.}

{Through the analysis performed in \cite{Mussi2018}, the parameter values summarized in Table \ref{tab:parameters} have been adopted for the aMGS in the reminder of this work. For the MGS-R, the following parameters have been adopted: $q=1/2K$, $\mathcal{I}=8K\sqrt{M}$, $R_{\rm max}=50$, $c_1=10$ and $c_2=0.5$ \cite{DattaSept.2013}.}

{\renewcommand{\arraystretch}{1.3}%
	\begin{table}[!htbp]
		\caption{{Best parameters for aMGS-MR detector presented in \cite{Mussi2018}}}
		{
			\begin{center}
				\small
				\begin{tabular}{rr|cccc}
					\hline
					& & \multicolumn{4}{c}{\bf \# aMGS samples, $L_e$}\\
					\textbf{Parameter} &  BS antennas & $\mathbf{1}$ &  $\mathbf{2}$&  $\mathbf{4}$ &  $\mathbf{8}$  \\ 
					\hline\hline 
					\multicolumn{1}{r|}{	\textbf{Mixing}}  & $N<=64$: &  $\frac{1}{4K}$  &  $\frac{1}{4K}$  &  $\frac{1}{3K}$&$\frac{1}{2K}$ \\
					\multicolumn{1}{r|}{\bf ratio, $q$} & $N>64$: & $\frac{1}{4K}$  &  $\frac{1}{4K}$ &  $\frac{1}{2K}$ & $\frac{1}{2K}$\\
					\hline
					\multicolumn{2}{r|}{\textbf{Max. \# Iterations, $I$}} & \multicolumn{4}{c}{{$3000$}} \\
					\hline
					\multicolumn{2}{r|}{\textbf{Max. Restarts}, $R_{\mathrm{max}}$} &  \multicolumn{4}{c}{{$5$}} \\
					\hline
					\multicolumn{2}{r|}{\textbf{Iterations stop criterion}, $c_1$ } & \multicolumn{4}{c}{{$10$}} \\
					\multicolumn{2}{r|}{\textbf{Restarts stop criterion}, $c_2$ } & \multicolumn{4}{c}{{$1$}} \\
					\hline
				\end{tabular}
			\end{center}
			\label{tab:parameters}
		}
\end{table}}

\begin{figure}[!htbp]
	\centering
	{\includegraphics[width=.65\textwidth]{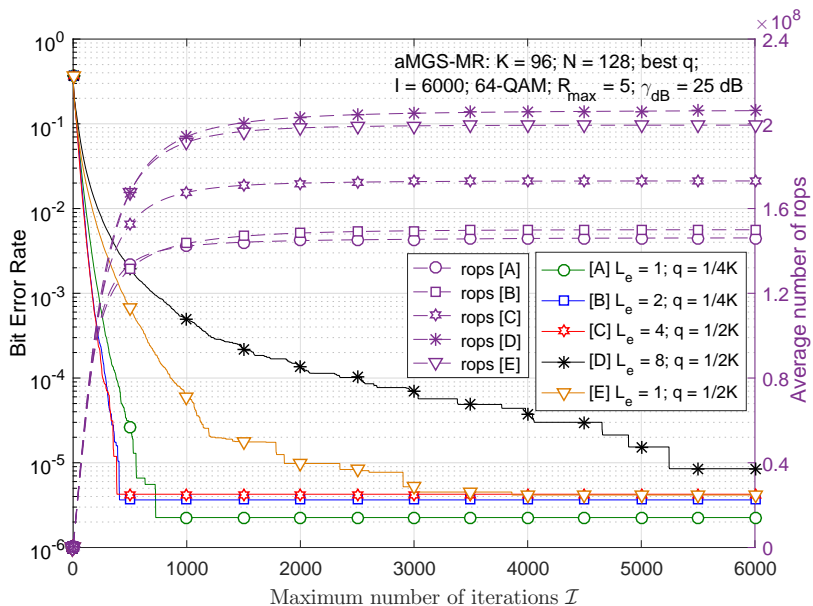}}
	\vspace{-3mm}
	\caption{BER performance and complexity \textit{vs.} convergence for the aMGS algorithm considering $64$-QAM modulation, best mixing ratio $q$ (curves [A] to [D]) from Table \ref{tab:parameters} and optimal value as proposed in \cite{DattaSept.2013}, i.e., $q=1/2K$ and $L_e=1$ (curve [E]) \cite{Mussi2018}.}
	\label{fig:complexity_96x128}
\end{figure}

{In Fig. \ref{fig:complexity_96x128} the convergence of the aMGS algorithm adopting best $q$ values, from Table \ref{tab:parameters}, is analysed against the average {\it rops} complexity, with $96\times128$ antennas and $64$-QAM \cite{Mussi2018}. For comparison purpose, a single sampling result using the optimal mixing ratio value as proposed in \cite{DattaSept.2013}, {i.e.}, $q=1/2K$ (curve [E]) is also included. One can notice that a less number of samples {has shown} to be {beneficial} in this LS-MIMO scenario, since the single {sample} case presented the best performance combined to the lowest asymptotic complexity, followed by the two ($ L_e = 2 $) and four-fold ($ L_e = 4 $) sampling case. Nevertheless, due to a slightly convergence gain observed with $L_e=2$ samples, the tradeoff {metric for} $L_e=1$ is found to be $ { \left.\chi\right|_{L_e=1} =39.83}$ against $ {\left.\chi\right|_{L_e=2} = 44.22}$ with $L_e=2$ samples. A detailed analysis of the aMGS performance/complexity gain in relation to the mixing ratio and the number of samples can be found in \cite{Mussi2018}. An in-depth analysis of the performance-complexity tradeoff of aMGS can be found in \cite{Mussi2018}.}

\subsection{Analysis on the proposed $d$-sMGS}\label{subsec:d_smgs_results}

\begin{figure} 
\centering
{\includegraphics[width=.65\textwidth]{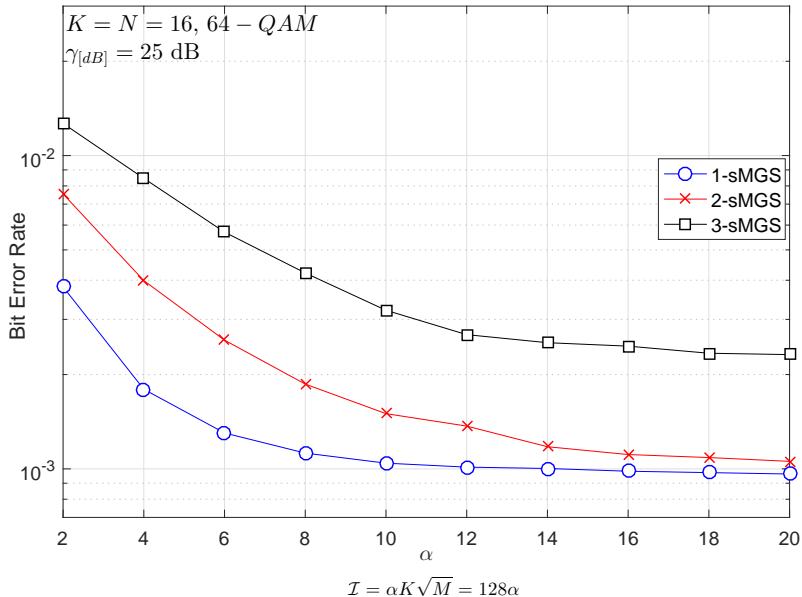}}
	\caption{{Performance convergence against the maximum number of iterations $\mathcal{I}$ of the proposed $d$-sMGS detector, with $16\times16$ antennas in 64-QAM modulation among different NL distance values. The number of iterations is related to the parameter $a$, were $\mathcal{I}=a K\sqrt{M}$.} } 
	\label{fig:convergence}
\end{figure}

{First of all, we focus on finding the maximum number of iterations $\mathcal{I}$ aiming at maximizing tradeoff performance $x$ complexity. In the literature, the quantity $\mathcal{I}=8K\sqrt{M}$ adopted in \cite{DattaSept.2013} is quite reasonable since it takes into account the number of active users and the modulation order. In this sense, Figure \ref{fig:convergence} shows the performance convergence of the proposed algorithm with the increase of the maximum number of iterations. We considered $K = N = 16$ antennas in 64-QAM with NL distance $d = \left\{1,2,3\right\}$ and used the parameter $a$ to denote the maximum number of iterations, so that $\mathcal{I}=aK\sqrt{M}=128a$. It can be clearly seen that the increase in the NL distance is not beneficial to the algorithm's performance, which is easily explained by the fact that, with increasing $d$, the neighborhood of the random solution increases, approaching the condition of unrestricted solution in the constellation, retaking its negative impact on the algorithm's convergence. Thus, observing the 1-sMGS curve, it can be seen that its convergence is reached with $a$ equal to $8$, which coincides with the result adopted in \cite{DattaSept.2013}. Therefore, this value $\mathcal{I}=8K\sqrt{M}$ will be adopted for the proposed $d$-sMGS detector in the reminder of this work.}

\begin{figure} 
	\centering
	\begin{subfigure}[$58\times64$]{\includegraphics[width=.49\textwidth]{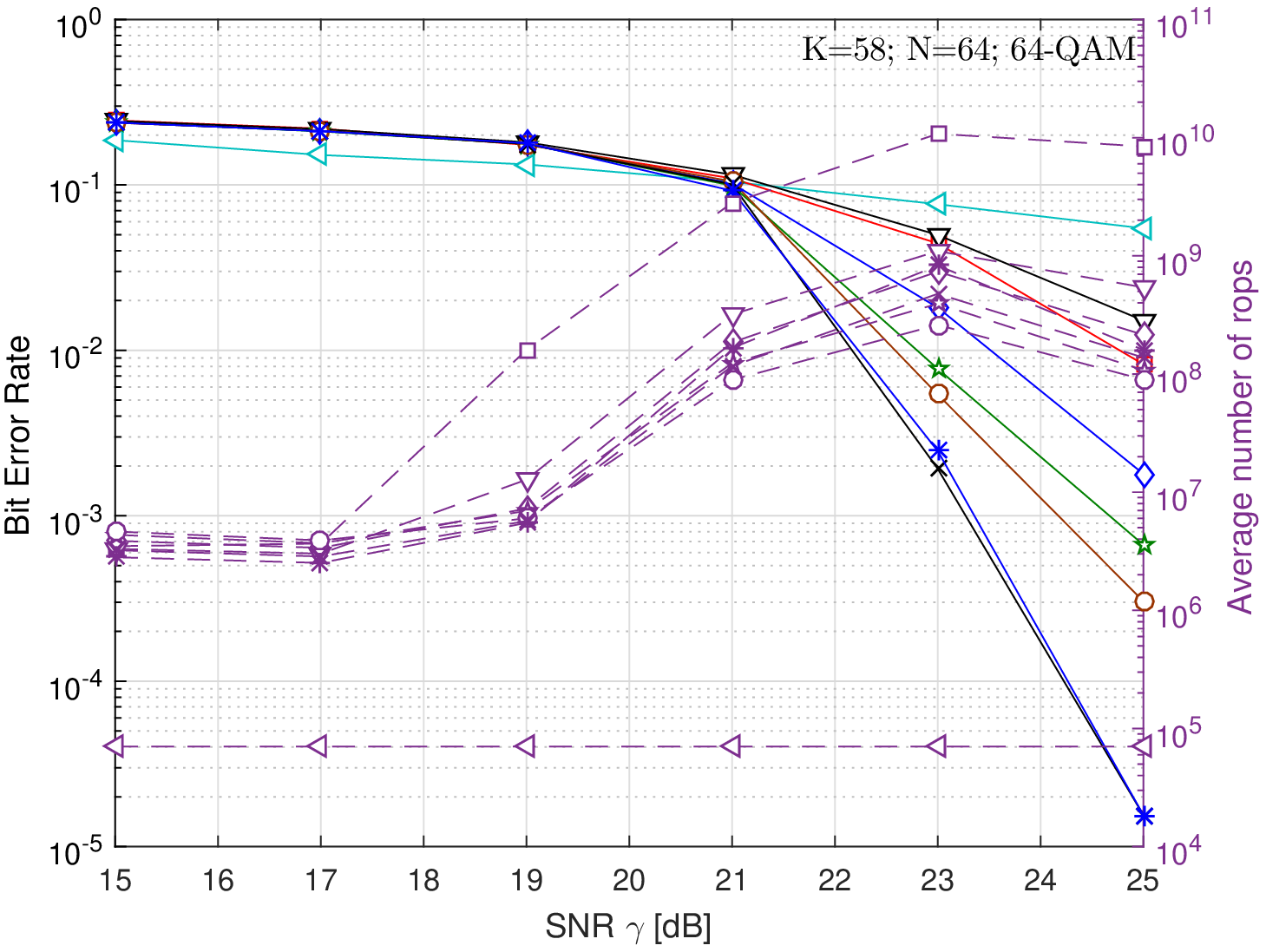}}
	\end{subfigure}
	\begin{subfigure}[$87\times96$]{\includegraphics[width=.49\textwidth]{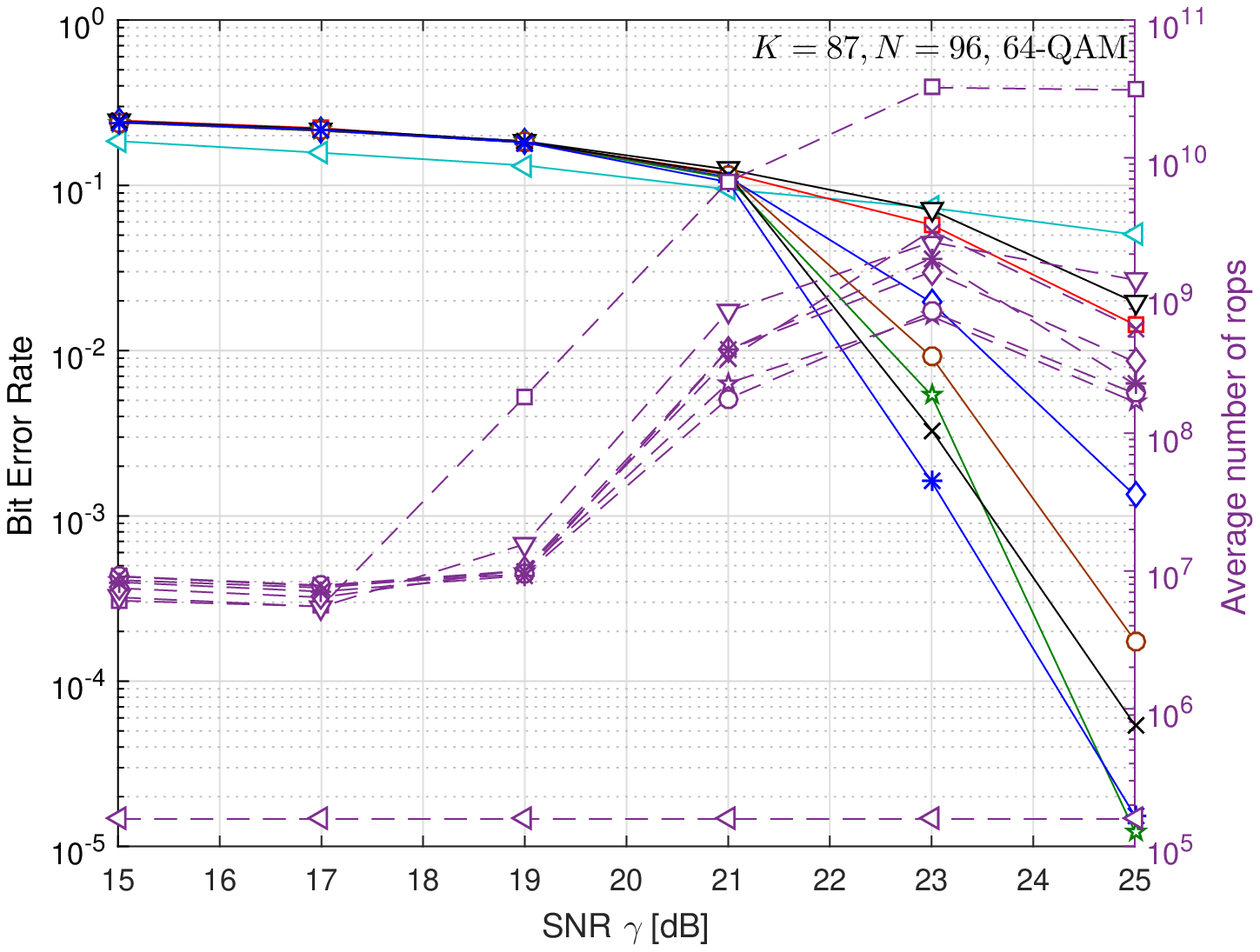}}
	\end{subfigure}
	\begin{subfigure}{\includegraphics[width=.4\textwidth]{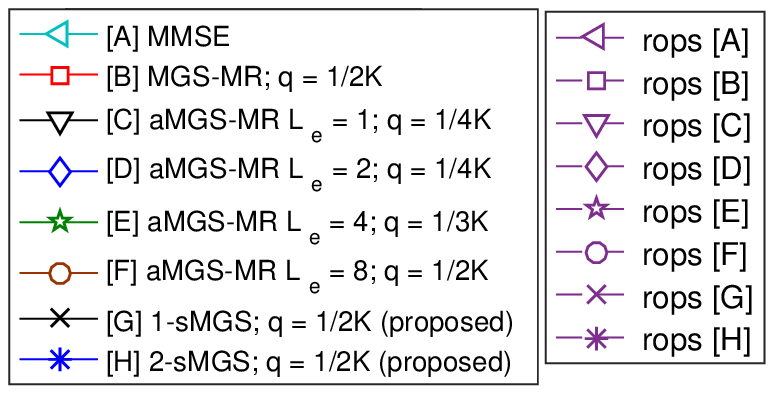}}
\end{subfigure}
	\caption{{SNR \textit{vs.} BER performance/Average number of {\it rops} in $d$-sMGS-MR detector against aMGS-MR approach and MGS-MR. Parameters: $K=58$, $N=64$, 64-QAM, $\beta \approx 0.9$.}}
	\label{fig:SNR_64QAM}
\end{figure}

{The Fig. \ref{fig:SNR_64QAM} shows the SNR \textit{vs.} performance - computational complexity of the addressed detectors. A high system loading, i.e., $\beta \approx 0.9$, in 64-QAM modulation is adopted with: a) $K = 58$, $N = 64$ and b) $K = 87$, $N = 96$ antennas. The parameters used for the MGS-MR and aMGS-MR detectors follow in their respective works: for the MGS-MR $\mathcal{I} = 8K\sqrt{M}$ and $R_{\rm max} = 50$ \cite{DattaSept.2013}; for the aMGS-MR $\mathcal{I} = 3000$, $R_{\rm max} = 5$ and the choice of the mixing ratio value is given according to the best option criterion published by the author \cite{Mussi2018}. One can notice in Fig. \ref{fig:SNR_64QAM}.a that both proposed detectors presented significant performance gain in the region of high SNR in relation to the other detectors, equivalent to approximately one decade against the second best performance detector aMGS-MR with $L_e = 8$ samples. Differently from that observed previously, the increase in the NL distance did not cause a loss of performance, since the $2$-sMGS detector resulted in a marginally similar performance to the $1$-sMGS. Thus, it denotes a tendency that the increase of the NL distance can be beneficial in scenarios with greater number of antennas, such as LS-MIMO. Related to the computational complexity, it can be observed that the complexity of the $1$-sMGS, $2$-sMGS and aMGS detectors with $L_e = 2$, $4$ and $8$ samples are marginally equivalent, although the aMGS with $8$ samples presented the least number of \textit{rops} (excluding the linear MMSE detector). Considering that both $d$-sMGS and aMGS have marginally the same complexity per iteration, it is shown that the strategy of multiple samples converged with fewer iterations, on the other hand, with inferior performance to that reached by $d$-sMGS.}

{With increasing antenna numbers, Fig. \ref{fig:SNR_64QAM}.b, it is reiterated the hypothesis that the increase of the NL distance results in a performance gain. One can notice a significant performance gain in the 4-sample aMGS detector, surpassing the result with $L_e = 8$, which corroborates the hypothesis that a smaller restriction in the noisy solution becomes beneficial with the increase in the number of antennas. In fact, in the region of high SNR, $\gamma_{\mathrm{dB}} = 25$ dB, it can be seen that the $2$-sMGS and aMGS with $L_e = 4$ achieve similar performance, although in the medium SNR region  ($\gamma_{\mathrm{dB}} = 23$ dB), the proposed $d$-sMGS still appear superior.
With respect to the complexity in terms of \textit{rops}, it is noticed that the $2$-sMGS-MR and aMGS-MR detectors with $L_e=4$ and $8$ samples presented a marginally equal complexity in $\gamma_{\mathrm{dB}} = 25$ dB; however, the least complexity is again reached by the aMGS, specially in medium SNR region ($\gamma_{\mathrm{dB}} = \left[21,23\right]$ dB). Therefore, it can be concluded that the proposed $d$-sMGS detection technique presented the best performance in both scenarios, and the smaller restriction of neighborhood with $d = 2$ was a more interesting choice with increasing number of antennas, in addition, there was no significant increase of complexity compared to the multiple sample detector aMGS, in other words, the complexity of the $2$-sMGS detector was marginally equal to the lowest complexity techniques: aMGS with $L_e = 4$ and $8$ samples.}

\begin{figure*}[!htbp]
	\centering	
	\begin{subfigure}[Performance, $N=64$]{\includegraphics[width=.49\textwidth]{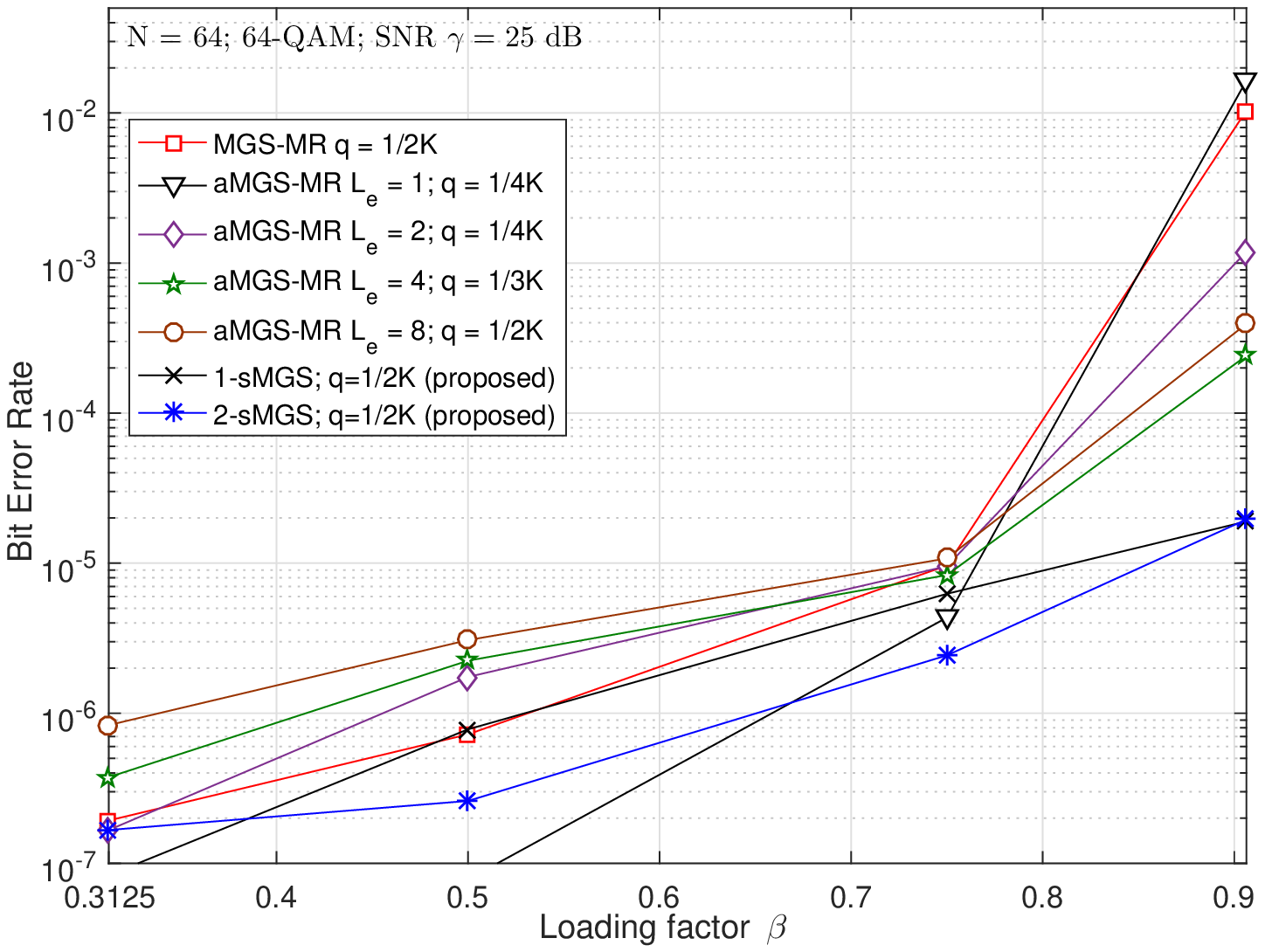}}
	\end{subfigure}
	\begin{subfigure}[Performance, $N=128$]{\includegraphics[width=.49\textwidth]{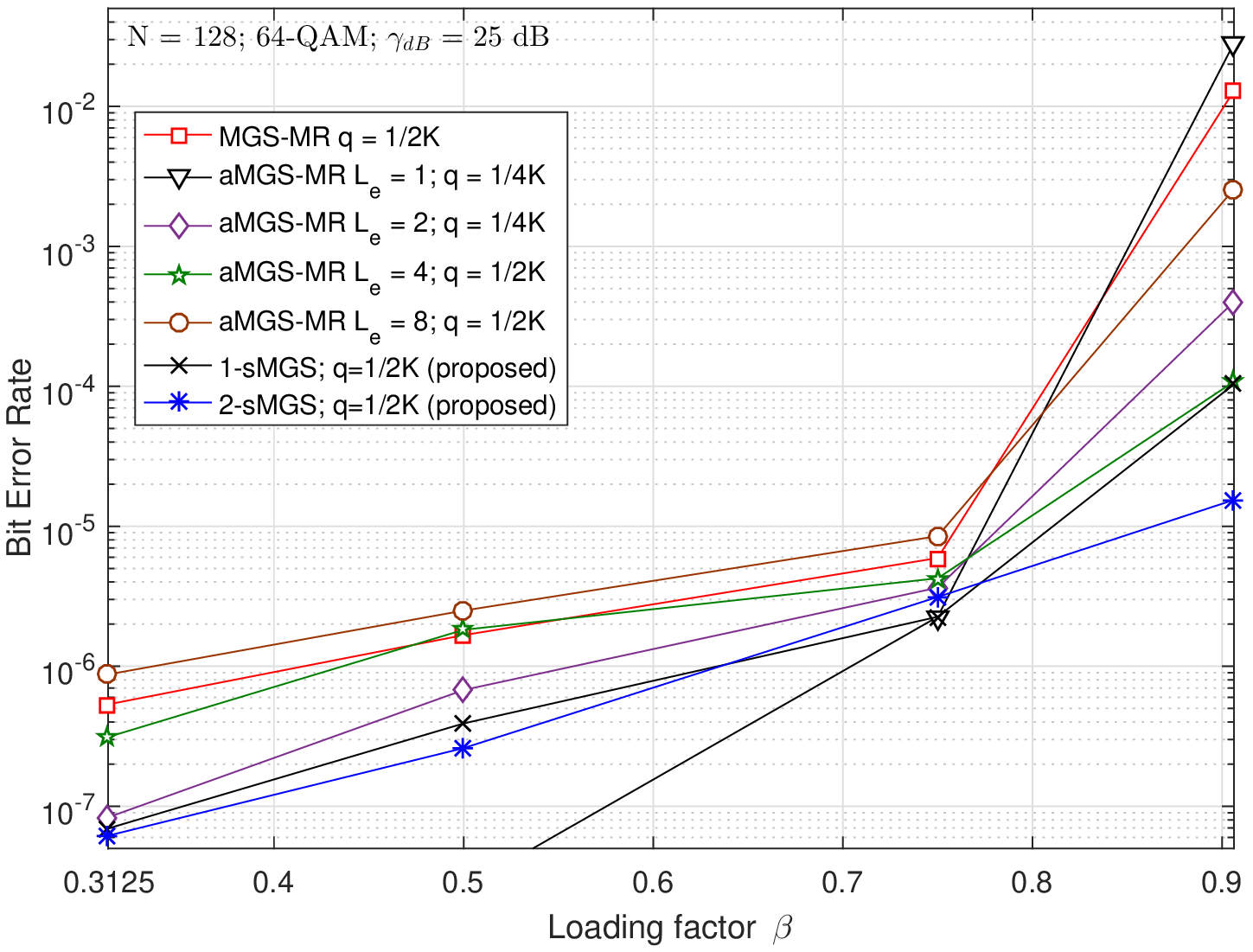}}
	\end{subfigure}
	\begin{subfigure}[Complexity, $N=64$]{\includegraphics[width=.49\textwidth]{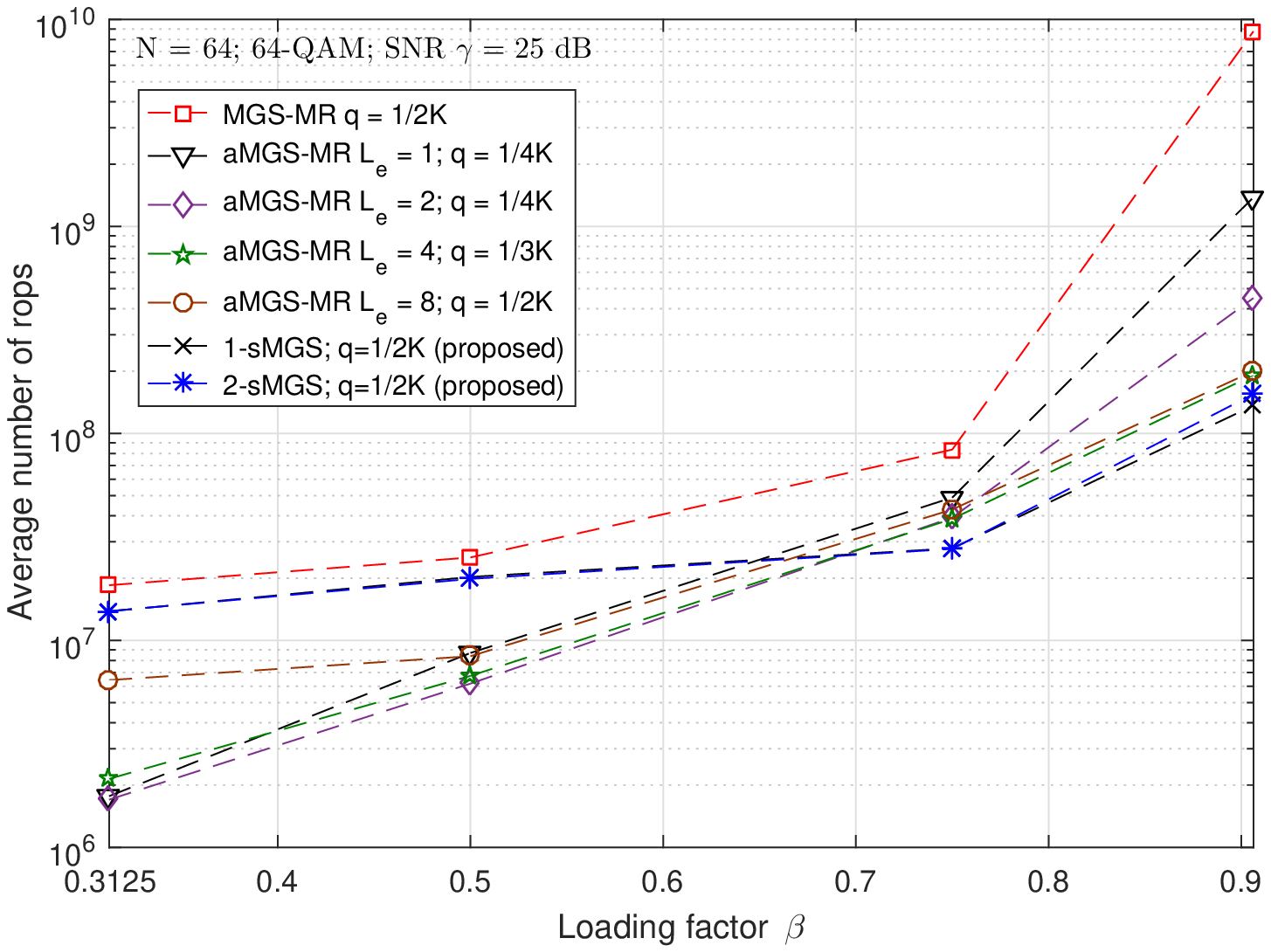}}
	\end{subfigure}
	\begin{subfigure}[Complexity, $N=128$]{\includegraphics[width=.49\textwidth]{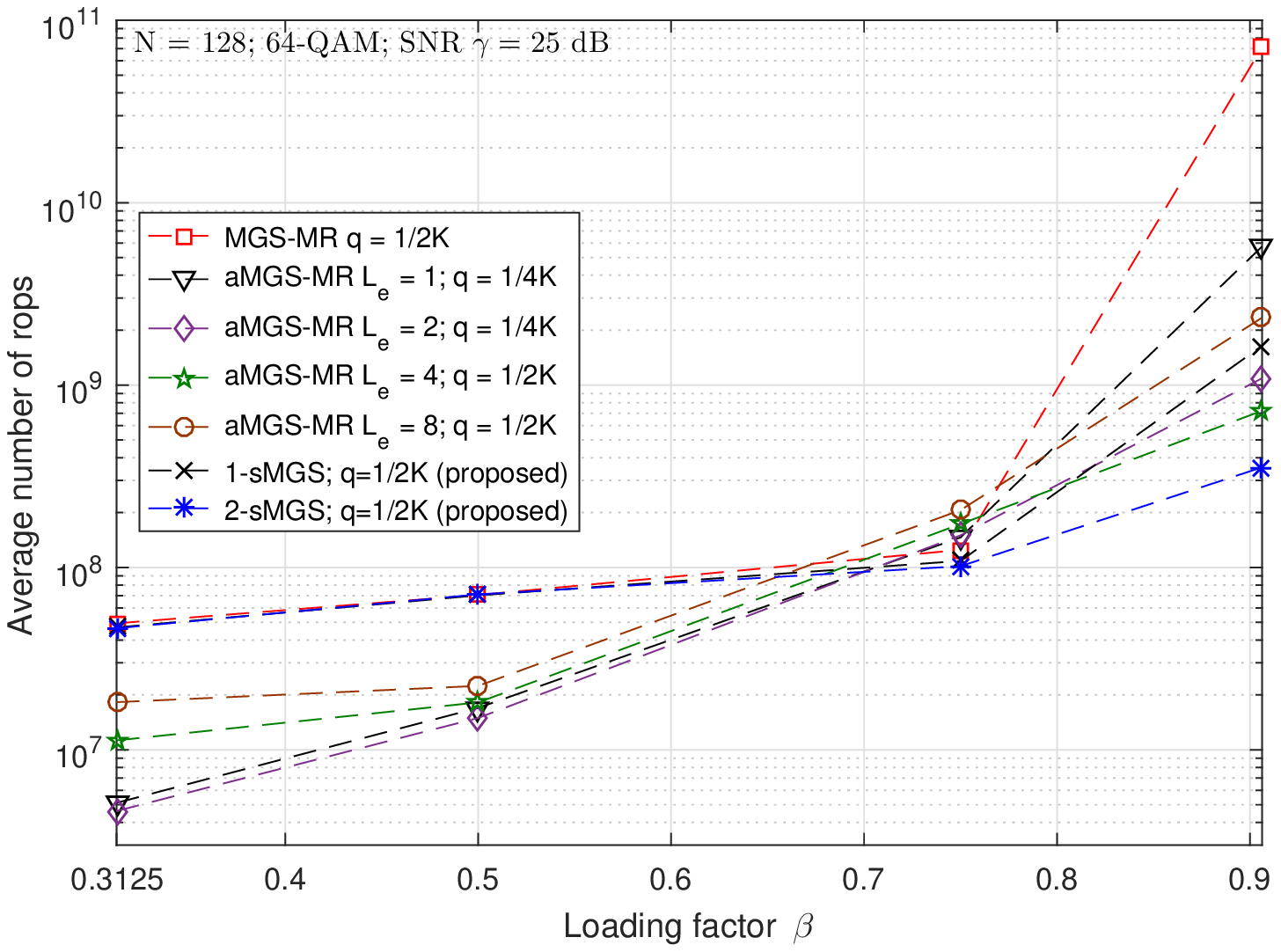}}
	\end{subfigure}
	\caption{Performance and complexity {\it versus} system loading, considering $\gamma_{\mathrm{dB}}=25$dB, $64$-QAM: {\bf a)} Performance for $N=64$;  {\bf b)} Performance for $N=128$ antennas; {\bf c)} Average {\it rop} complexity for $N=64$; {\bf d)} Average {\it rop} complexity for $N=128$.} \label{fig:LF_64QAM_25dB}
\end{figure*}

{A system loading analysis against BER and  {\it rops} complexity is depicted in Fig. \ref{fig:LF_64QAM_25dB} under $\gamma_{\mathrm{dB}}=25$ dB. It may be first noted that at high loading, i.e., $\beta \approx 0.9$, the proposed detection scheme showed a significant gain in performance over the aMGS. In the other regions, there is no clearly  outstanding technique, however, a lower restriction in the noisy solution demonstrated better results, which are represented by the $2$-sMGS overpassing the $1$-sMGS and aMGS with $L_e = 1$ or $2$ in front of the $L_e = 4$ and $8$ samples. In relation to the computational complexity with $N = 64$ antennas (Fig. \ref{fig:LF_64QAM_25dB}.b), one can notice that in the medium-high loading region ($\beta \geq 0.75$), the proposed $d$-sMGS strategy presented less complexity both with respect to multiple sampling aMGS and conventional MGS. In the medium-low system loading results ($\beta \leq 0.5$), multiple sampling schemes presented lower computational complexity. Therefore, one can highlight the superiority of the proposed strategy in both performance and complexity in medium-high loading configurations, demonstrating the potential of this strategy when the LS-MIMO system operates under high loading crowded scenarios. This can be explained as the number of mobile users increases, approaching the full-loading system condition $\beta \rightarrow 1$, the set of possible symbol combinations becomes larger, such that the noisy solution from the mixture has its negative effect aggravated, affecting the algorithm's convergence; whereas the NL strategy is able to mitigate this effect, having a beneficial effect on the convergence which results in improvement in performance and complexity reduction.}

{With the increasing number of antennas at $N = 128$, the system loading analysis reflects a clear superiority of the $2$-sMGS detector in high loading configurations, both in performance and in complexity. This performance behavior corroborates the hypotheses raised in Fig. \ref{fig:SNR_64QAM} regarding performance improvement with increasing NL distance. On the other hand, in medium-low loading, the complexity of $2$-sMGS was shown to be greater than aMGS and $1$-sMGS, equating only to the conventional MGS-MR.}

\section{Conclusions}\label{sec:concl}
{A neighborhood limited $d$-sMGS detector for large-scale MIMO systems has been proposed based on the neighborhood constraint of the noisy solution at a distance of $d$.}

{The proposed LS-MIMO $d$-sMGS detection scheme demonstrated the ability to mitigate the impact caused by the noisy solution from the mixture, which is aggravated and can become harmful when the full system loading condition is present or when a high order modulation is implemented.}

{The modifications in the MGS technique proposed here have demonstrated effectiveness in achieving convergence improvements in the detection algorithm, which resulted in significant gains in performance and complexity compared to both the multiple sampling aMGS technique as well as the conventional MGS. These advantages are especially obtained when the system loading is high and there are a large number of antennas, condition favorable to LS-MIMO. Moreover, with increasing the number of dimensions, i.e., increasing number of antennas and/or modulation order, a smaller restriction of $2$-sMGS was shown to be a more interesting choice than $1$-sMGS.}
	
{In addition, the NL strategy represented less complexity per iteration compared to aMGS or MGS, since only one sample is calculated and the simplified objective function is considered. On the other hand, when a low system loading is considered, the NL strategy resulted in a slight increase in complexity.}

\section*{Abbreviations}
aMGS: averaged MGS; BS: base station; BER: bit error rate; {$d$-sMGS: $d$-simplified MGS;} ENI: effective number of iterations; LS-MIMO: large-scale multiple-input multiple-output; MCMC: Markov chain Monte Carlo; MMSE: minimum mean square error; MGS: mixed Gibbs sampling; MR: multiple restart; MS: multiple sampling; {NL: neighborhood limitation}; SNR: signal-to-noise ratio. 
\section*{Acknowledgements}
We gratefully acknowledge the agencies: National Council for Scientific and Technological Development (CNPq) of Brazil, the University of São Paulo (USP), the State University of Londrina (UEL), the Federal Institute of Paraná (IFPR) and the Paraná State Government. 
This work has been partially supported by the CNPq of Brazil under Grants 304066/2015-0, by the USP, by the UEL, by the IFPR and the Paraná State Government.



\end{document}